\documentclass[floatfix,reprint,superscriptaddress,amsmath,amssymb,aps,prx,longbibliography]{revtex4-2}

\DeclareSymbolFont{usualmathcal}{OMS}{cmsy}{m}{n}
\DeclareSymbolFontAlphabet{\mathcal}{usualmathcal}

\usepackage{cancel}
\usepackage[normalem]{ulem}
\usepackage[export]{adjustbox}
\usepackage{amsmath}
\usepackage{amsfonts}
\usepackage{amssymb}
\usepackage{mathtools}
\usepackage{graphicx}
\graphicspath{{./imagesOperator} }
\usepackage{physics}
\usepackage{microtype}
\usepackage{verbatim}
\usepackage{bbold}
\usepackage[colorlinks]{hyperref}
\usepackage{cleveref}
\allowdisplaybreaks
\usepackage{multirow}
\usepackage{bm}
\usepackage{bbm}
\usepackage{graphicx}
\usepackage{braket}
\usepackage{todonotes}
\usepackage{hyperref}
\usepackage{layouts}
\usepackage{color,soul}
\usepackage{array}
\usepackage{float}
\usepackage{algorithm2e}
\usepackage{algpseudocode}

\newcommand{\e}{\mathrm{e}}
\newcommand{\rd}{\mathrm{d}}
\newcommand{\I}{\mathrm{i}}

\makeatletter
\renewcommand{\Indentp}[1]{
  \advance\leftskip by #1
  \advance\skiptext by -#1
  \advance\skiprule by #1}
\renewcommand{\Indp}{\algocf@adjustskipindent\Indentp{\algoskipindent}}

\renewcommand{\Indm}{\algocf@adjustskipindent\Indentp{-\algoskipindent}}

\makeatother

\graphicspath{{./imagesOperator/}}
\definecolor{Ured}{HTML}{cc0000}
\definecolor{Ublue}{HTML}{1f65cf}
\definecolor{Ugreen}{HTML}{198a11}

\begin{document}
	\title{Predicting Dynamics from Flows of the Eigenstate Thermalization Hypothesis}
	
	\author{Dominik Hahn}
    \email{dominik.hahn@physics.ox.ac.uk}
    \affiliation{Rudolf Peierls Centre for Theoretical Physics, Clarendon Laboratory, Oxford OX1 3PU, UK}
    \affiliation{Max Planck Institute for the Physics of Complex Systems, 01187 Dresden, Germany}
    \author{David M. Long}
    \affiliation{Condensed Matter Theory Center and Joint Quantum Institute,\\Department of Physics, University of Maryland, College Park, Maryland 20742, USA}
    \affiliation{Department of Physics, Stanford University, Stanford, California 94305, USA}
    \author{Marin Bukov}
    \affiliation{Max Planck Institute for the Physics of Complex Systems, 01187 Dresden, Germany}
    \author{Anushya Chandran}
    \affiliation{Max Planck Institute for the Physics of Complex Systems, 01187 Dresden, Germany}
    \affiliation{Department of Physics, Boston University, Boston, Massachusetts 02215, USA}

\begin{abstract}
Analytical treatments of far-from-equilibrium quantum dynamics are few, even in well-thermalizing systems.
The celebrated eigenstate thermalization hypothesis (ETH) provides a post hoc ansatz for the matrix elements of observables in the eigenbasis of a thermalizing Hamiltonian, given various response functions of those observables as input.
However, the ETH cannot predict these response functions.
We introduce a procedure, dubbed the statistical Jacobi approximation (SJA), to update the ETH ansatz after a perturbation to the Hamiltonian and predict perturbed response functions.
The Jacobi algorithm diagonalizes the perturbation through a sequence of two-level rotations.
The SJA implements these rotations statistically assuming the ETH throughout the diagonalization procedure, and generates integrodifferential flow equations for various form factors in the ETH ansatz.
We approximately solve these flow equations using statistical information about Jacobi rotations as input, and predict both quench dynamics and autocorrelators in the thermal state of the perturbed Hamiltonian.
The predicted dynamics agree with exact numerics in both random matrix models and spin chains, outperforming second-order time-dependent perturbation theory.
\end{abstract}

\maketitle
\section{Introduction}\label{sec:intro}

It is difficult to make quantitative \emph{predictions} of the dynamics of an isolated quantum many-body system, even in well-thermalizing systems. However, there is a successful \emph{description} of observables in such systems, given various response functions for those observables as input; this is the eigenstate thermalization hypothesis (ETH)~\cite{Jensen1985statistical,Deutsch1991ETH,Srednicki1994ETH,Rigol2008ETHnumerics,DAlessio2016ETHreview,Deutsch2018ETHreview,buvca2023unified,doyon2017thermalization,helbig2024theoryeigenstatethermalisation}. Extending seminal random matrix models of complex atomic nuclei~\cite{Wigner1955rmt}, the ETH was formulated in the 1980's~\cite{Jensen1985statistical}, and significantly generalized in the 1990's~\cite{Deutsch1991ETH,Srednicki1994ETH}. 
It hypothesizes that \emph{individual eigenstates are thermal}. That is, few-body expectation values and correlation functions in individual eigenstates are the same as those in the thermal ensemble. 
This hypothesis has been tested numerically using exact diagonalization~\cite{Rigol2008ETHnumerics,Beugeling2014Finite,DAlessio2016ETHreview,SteiningwegEigenstate2013,Kim2014testing,Chandran_2016eigenstate,Steiningweg2014Pushing,Khatami2012QuantumQuenches,KhodjaRelevance2015,Beugeling2015Offdiagonalmatrix,Genway2012Thermalization,Birolo2010Effect,Roux2010Finite,Sorg2014Relaxation}. Conversely, non-ergodic behavior can be detected from eigenstates as a violation of ETH, as exhibited by quantum many-body scars~\cite{Turner_2018,Chandran_2023}, Hilbert space fragmentation~\cite{Sala2020,Moudgalya_2022}, and many-body localization~\cite{Nandkishore_2015,Abanin_2019Colloquium}.

A different perspective on the ETH is that it is a maximal entropy ansatz for the matrix elements of a local operator in the energy eigenbasis of a thermalizing system, subject to the constraint of reproducing response and correlation functions of interest. For the matrix elements of an operator $A$ in the energy eigenbasis $|i_0\rangle $ of a Hamiltonian $H_0$, it reads,
\begin{align}
\braket{i_0|A|j_0}=
A(E)\delta_{i_0,j_0}+
\frac{f_A(E,\omega)}
{\sqrt{\nu(E+\omega)}}R_{i_0j_0},\label{Eq:ETHOffDiag}
\end{align}
where $H_0\ket{i_0}=E_{i_0}\ket{i_0}$,
$E=E_{i_0}$, and $\omega=E_{j_0}-E_{i_0}$ is the energy difference, $A(\overline{E})$ and $f_A(E, \omega)$ are smooth functions of their arguments (dubbed form factors), $\nu(E)$ is the density of states at energy $E$, and $R_{i_0j_0}$ are pseudo-random variables with mean zero and variance one. 

 The energy-dependent \emph{form factors}, $A(E)$ and $f_A(E,\omega)$, ensure that individual eigenstates reproduce the microcanonical expectation value and autocorrelator of $A$ in the thermodynamic limit~\cite{srednicki1999approach}.

\begin{figure}[t!]
        \centering
        \includegraphics[width=0.4\textwidth]{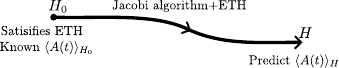}
        \caption{\emph{The statistical Jacobi approximation}~(SJA) predicts the dynamics of a well-thermalizing Hamiltonian $H$, given a Hamiltonian $H_0$ satisfying the ETH, observable dynamics generated by the Hamiltonian $H_0$, denoted by $\braket{A(t)}_{H_0}$, and statistical information about the Jacobi algorithm. Specifically, it takes the form factors appearing in the ETH ansatz for the Hamiltonian $H_0$ [Eq.~\eqref{Eq:Spectralfunctions}] and a joint number density [Eq.~\eqref{eq:rhodecdefjoint}] as input, and computes the form factors for the Hamiltonian $H=H_0+JV$. Response functions of the Hamiltonian $H$, e.g., $\braket{A(t)}_{H}$, follow from the form factors.
        } 
        \label{fig:sketchProcedure}
\end{figure}

In the last decade, several practitioners have extended the ETH to describe multi-point correlation functions, out-of-time ordered correlators~\cite{Pappalardi2022Eigenstate,Foini2019Eigenstate,Hahn2024Eigenstate,Chan2019Eigenstate} and responses to sudden changes in parameters (quenches)~\cite{Jindal_2024,foini2024outofequilibriumeigenstatethermalizationhypothesis}. These extensions continue to be maximal entropy ansatze for matrix elements of operators as in Eq.~\eqref{Eq:ETHOffDiag}, but with cross-correlations between the pseudo-random numbers $R_{ij}$ (in the same matrix or between matrices) to encode the desired response functions.

Going from a descriptive hypothesis to a predictive theory of quantum dynamics requires calculating the various energy-dependent form factors in the ETH ansatz---$A(E)$ and $f_A(E, \omega)$ in Eq.~\eqref{Eq:ETHOffDiag}. This article takes the first steps in this direction. We assume the validity of ETH in the eigenbasis of a given Hamiltonian $H_0$ and in a rotated basis produced by the Jacobi algorithm~\cite{Jacobi1846,Golub2000} to derive ETH ansatze  [Eq.~\eqref{Eq:ETHOffDiag}] with respect to a target Hamiltonian $H$. These ansatze predict form factors, and thus various response functions, with respect to the Hamiltonian $H$, given form factors with respect to the Hamiltonian $H_0$ and quantitative statistical information about the Jacobi algorithm as input.

The Jacobi algorithm is a numerical algorithm to diagonalize a matrix using a series of two-level unitary
rotations. Each two-level rotation is designed to zero out the two largest in magnitude off-diagonal matrix elements, denoted by \(w\). When applied to the matrix $H_{i_0j_0}$, it produces a series of two-level rotations that rotate the eigenbasis of $H_0$ into the eigenbasis of $H$~[Fig.~\ref{fig:sketchJacobi0}].
The pseudo-randomness of the matrix $A$ in the eigenbasis of $H_0$ [Eq.~\eqref{Eq:ETHOffDiag}] suggests that these rotations can be performed statistically and independently, at least until a typical row/column of the starting matrix has been significantly updated by the algorithm. Let the number of these iterations be $n$. We posit the ETH after $n$ iterations; i.e., the matrix elements of \(A\) in the rotated basis read,
\begin{equation}\label{Eq:Spectralfunctions}
A^{(n)}_{ij}=A^{(n)}(E)\delta_{ij}+\frac{f^{(n)}_A(E,\omega)}{\sqrt{\nu(E+\omega)}}R^{(n)}_{ij}.
\end{equation}

Given the statistics of the $n$ rotation angles, the form factors $A^{(n)}(E)$ and \(f_A^{(n)}(E,\omega)\)  are linearly related to the given form factors in Eq.~\eqref{Eq:ETHOffDiag}. 
We keep going, performing another $n$ iterations, and positing the ETH again. This entire procedure, called the \emph{statistical Jacobi approximation}~(SJA)~\cite{Long2022phenomenology,Long2023Beyond}, produces an integrodifferential flow equation for the form factors [Eq.~\eqref{Eq:resultshort}] \emph{in the thermodynamic limit}, as a function of the current largest matrix element \(w\). The form factors at \(w=0\) determine the post-quench dynamics, and themselves are determined by the statistical distribution of Jacobi rotations and the form factors at $n=0$~[Fig.~\ref{fig:sketchProcedure}]. 

We derive flow equations for form factors that predict autocorrelators \(\braket{A(t)A(0)}_{H}\) in the thermal ensemble of the perturbed Hamiltonian $H$, and quench dynamics \(\braket{A(t)}_H\) upon quenching from $H_0$ to $H$. As we cannot solve the flow equations exactly, we solve them iteratively, and produce solutions as a series expansion in $J$, where $H-H_0 \eqqcolon JV$. The predicted dynamics of $A$ quantitatively agrees with exact numerics in random matrix models and in one-dimensional spin chains, and is more accurate than second-order time-dependent perturbation theory. The latter result deserves emphasis: despite perturbatively solving the flow equations, the predicted dynamics is non-perturbative in $J$ because of the statistical input from the Jacobi algorithm.

The flow equations depend on a single statistical and non-perturbative input from the Jacobi algorithm, which is a joint number density~[cf. Eq.~\eqref{eq:rhodecdef} and Eq.~\eqref{eq:rhodecdefjoint}] of: the matrix element \(w\) that is zeroed out; the eigenstate energies \(E, E’\) connected by this element; and the matrix element of \(A\) between those eigenstates. Beyond that, we only require the initial form factors.
We obtain the joint number density and initial conditions from small-system numerics.
In well-thermalizing systems with short correlation lengths, this distribution and the initial conditions remain stable as the system size increases. Intuitively, this is because the Jacobi algorithm rotates states most strongly coupled by the perturbation $JV$ first. These rotations have the largest ability to rearrange the spectrum of $A$, and thus the largest impact on dynamical responses. 

The SJA has been previously used by a subset of the authors in a few contexts. In the setup described above, the SJA provides a closed form solution to the (log) survival probability of an eigenstate of $H_0$ after a quench from $H_0$ to $H$~\cite{Long2023Beyond}. The solution is quantitatively accurate, and captures corrections to the Fermi Golden Rule rate of decay for large perturbations. 
However, the survival probability is not a very useful measure of thermalization in many-body systems, as it decays with an extensive rate, and becomes unmeasurable long before thermalization actually occurs. 
The extension of the SJA to the context of ETH allows us to probe observable physics.
The philosophy of the SJA was also previously applied to pre-thermal many-body localized systems~\cite{Long2022phenomenology}. Using a different statistical description of matrix elements in the eigenbasis of $H_0$, as opposed to the ETH, Ref.~\cite{Long2022phenomenology} predicted stretched exponential decay of local auto-correlators. Numerically exact calculations confirmed this prediction.

There are other approaches to iteratively diagonalize a matrix and derive flow equations for response functions, notably the Wegner-Wilson flow~\cite{Wegner1994,kehrein2007flow,Moeckel2008Interaction,Moeckel_2009}. The chief divergence from the SJA is the applicability of the ETH during the flow. The Wegner-Wilson scheme, for instance, rotates the eigenbasis of $H_0$ to that of $H$ through a sequence of rotations that have spatially local generators, and thus intricate correlations in Hilbert space. In special cases where the observable is the perturbation~\cite{Richter_2019,Richter_2019b} or the perturbation is a random matrix~\cite{Dabelow_2020,Dabelow_2021,Belitti2019}, analytical results for the response functions can be obtained.

The paper is organized as follows.
Sec.~\ref{sec:Jacobi} reviews theoretical background on the Jacobi algorithm and the ETH. Sec.~\ref{sec:Stat JAcobi} introduces the SJA and flowing form factors.
Sec.~\ref{sec:flow} derives flow equations for the form factors and Sec.~\ref{sec:flow_sol} presents their iterative solution.  
Sec.~\ref{sec:numerics} explains the numerical implementation and compares these solutions to numerically exact computations in random matrix models and one-dimensional spin chain models.  
Finally, Sec.~\ref{sec:disc} closes with a discussion of extensions and applications.

\subsection{Summary of main result}\label{subsec:overview}
In this subsection, a high-level overview of the main technical result is presented. While similar results are obtained for autocorrelators, the focus here is on expectation values.
 
Given the dynamics of an observable under \(H_0\),
the objective is to predict $\braket{A(t)}_H$ with respect to the perturbed Hamiltonian $H=H_0+JV$.
We define the form factor $B(w_0, E,\omega)$ using the Lehmann representation in the eigenbasis of $H_0$,
\begin{align}\label{eq:Dynamics0}
\begin{split}
		\braket{A(t)}_{H_0}-\braket{A(0)}_{H_0} &= \int \mathrm{d}E\mathrm{d}\omega B(w_0, E,\omega) \left(\e^{\I \omega t}-1\right),
\end{split}
\end{align}
with 
\begin{align}\label{Eq:Spectralfunctions}
B(w_0, E,\omega)&=\frac{1}{\rd E\rd \omega}\sum_{\substack{E_{i_0} \in [E, E+\mathrm{d}E)\\E_{j_0} \in [E+\omega, E+\omega+\mathrm{d}\omega)\\i_0\neq j_0}}\braket{i_0|\rho|j_0} \braket{j_0|A|i_0}.
\end{align}
The initial density matrix at $t=0$, $\rho$, is stationary with respect to $H_0$, e.g. a microcanonical ensemble with uniform weight in a finite energy window. Here,
$w_0$ is the starting value of the flow parameter $w$ at Jacobi iteration \(n=0\), and is given by the largest off-diagonal matrix element, in absolute value, of the perturbation $J V$ in the eigenbasis of $H_0$. Form factors with respect to the Hamiltonian $H_0$ ($H$) correspond to flow parameters $w=w_0$ ($w=0$).

The Lehmann representation also connects the dynamics of the observable $A$ with respect to the Hamiltonian $H$ to the form factor with respect to the same Hamiltonian:
\begin{align}\label{eq:Dynamics0}
\begin{split}
		\braket{A(t)}_{H}-\braket{A(0)}_{H} &= \int \mathrm{d}E\mathrm{d}\omega B(0, E,\omega) \left(\e^{\I \omega t}-1\right),
\end{split}
\end{align}
where
\begin{align}
B(w=0, E,\omega)&=\frac{1}{\rd E\rd \omega}\sum_{\substack{E_{i} \in [E, E+\mathrm{d}E)\\E_{j} \in [E+\omega, E+\omega+\mathrm{d}\omega)\\i\neq j}}\braket{i|\rho|j} \braket{j|A|i} \label{Eq:Bw0def}, 
\end{align}
with $\ket{i},\,\ket{j}$ eigenstates of the Hamiltonian $H=H_0+JV$.

We derive an integrodifferential equation
 governing the evolution of $B(w, E,\omega)$ under Jacobi rotations in Sec.~\ref{sec:flow}, where \(B(w, E,\omega)\) is the form factor in the Jacobi-updated basis.
The derivation assumes the ETH all the way along the flow, which we expect to hold for thermalizing systems.

Solving the differential equation for $B(w, E, \omega)$ is difficult in general. In Sec.~\ref{sec:flow_sol}, we solve the equation by representing the solution as a series expansion in $J$.
For instance, in the case where $A$ and $H_0$ commute, to leading order in \(J\), the form factor at the end of the Jacobi flow is: 
\begin{align}\label{eq:premresult}
	B(w=0,E,\omega)&=J^2\frac{|f_{\mathrm{Jac}}(E,\omega)|^2}{\omega^2}\times	\nonumber\\&\Biggl\{\Bigl[p(w_0,E)-\frac{\nu(E)}{\nu(E+\omega)} p(w_0,E+\omega) \Bigr]\nonumber\\
    &\times\Bigl[A(w_0,E)-A(w_0,E+\omega)\Bigr] \Biggr\},
\end{align}
where $A(w_0,E)$ and $p(w_0,E)$ are determined by \(H_0\);
    	\begin{align}\label{eq:definitions}
        A(w_0,E)\mathrm dE&= \frac{1}{\nu(E)}\sum_{E_{i_0} \in [E, E+\mathrm{d}E]} \braket{i_0|A|i_0}\\
    p(w_0,E)&= \frac{1}{\mathrm{d} E}\sum_{E_{i_0} \in [E, E+\mathrm{d}E]} \braket{i_0|\rho|i_0}.
	\end{align}

The \emph{Jacobi spectral function}, $|f_{\mathrm{Jac}}(E,\omega)|^2$, is the statistical input from the Jacobi algorithm. It is the sum squared of the decimated elements that connect the energies $E$ and $E+\omega$. At lowest order in $J$, it agrees with the spectral function of the perturbation $f_{V}(E,\omega)$ with respect to the Hamiltonian $H_0$~\cite{Long2023Beyond}:
\begin{align}\label{eq:fJac}
	|f_{\mathrm{Jac}}(E,\omega)|^2=|f_{V}(E,\omega)|^2+\mathcal{O}(J^2),
\end{align}
so that Eq.~\eqref{eq:premresult} recovers second-order time-dependent perturbation theory. However, as $|f_{\mathrm{Jac}}(E,\omega)|^2$ is obtained by numerically implementing the Jacobi algorithm, it includes non-perturbative corrections in $J$. 

In Sec.~\ref{sec:numerics}, we compare the predictions of iterative approximations of the flow equations to second-order time-dependent perturbation theory, and find quantitatively more accurate predictions for autocorrelators and expectation values when we use the numerically exact $|f_{\mathrm{Jac}}(E,\omega)|^2$. We anticipate that solving the differential equation for $B(w,E,\omega)$ to higher orders in $J$ should increase accuracy, until some order where the asymptotic expansion ceases to be effective.

\section{Theoretical background}
\label{sec:Jacobi} 
    The following two subsections review the Jacobi diagonalization algorithm (Sec.~\ref{subsec:Jac_alg}) and the out-of-equilibrium ETH ansatz (Sec.~\ref{subsec:ETHspectralforms}). The out-of-equilibrium ETH ansatz is an extension of Eq.~\eqref{Eq:ETHOffDiag} that captures the entire dynamics of an operator $A$ after a quench to the Hamiltonian $H_0$. Sec.~\ref{subsec:ETHspectralforms} also explains how the form factors in the ETH ansatz encode different response functions. 

    \subsection{Jacobi diagonalization algorithm}
        \label{subsec:Jac_alg}
            \begin{figure*}[t]
        \centering
        \includegraphics[width=1\textwidth]{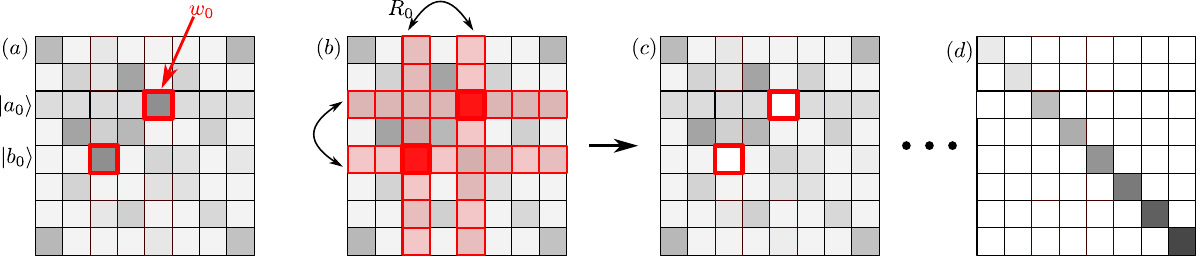}
        \caption{Sketch of the Jacobi algorithm: (a) At each iteration step, identify the largest off-diagonal matrix element $w_0$. This is the element to be decimated. (b) Perform the 2-level unitary rotation $R_0$ that sets $H_{a_0b_0}$ to zero. This rotation affects the rows and columns associated with indices $a_0$ and $b_0$. (c) After the rotation, $H_{a_1b_1}=0$. (d) Repeated decimations lead to a full diagonalization of the matrix in $\mathcal{O}(N^2)$ rotations, with $N$ being the matrix size. The grayscale denotes the absolute value of the matrix elements, increasing from white to black.
        } 
        \label{fig:sketchJacobi0}
        \end{figure*}
        The Jacobi diagonalization algorithm~\cite{Jacobi1846} is an iterative procedure to diagonalize an \(N\times N\) Hermitian matrix. 
        
        Consider a matrix \(H_{j_0 k_0}\) in an arbitrary computational basis \(\{\ket{j_0}\}\), for $j_0=1,\dots, N$. The algorithm diagonalizes the matrix by applying a sequence of two-level rotations, which \emph{decimate} the largest off-diagonal elements (see Fig.~\ref{fig:sketchJacobi0}). 
        It proceeds as follows:
        \begin{enumerate}
        
        \item\label{itm:algorithm} Find the largest, in absolute-value, off-diagonal matrix element,
        \begin{equation}
            w_0 = \max_{j_0 \neq k_0} |\!\braket{j_0|H|k_0}\!| = |\!\braket{a_0|H|b_0}\!|.
        \end{equation}
        The corresponding submatrix (with $a_0$ and $b_0$ as row-and column indices) is given by 
        \begin{equation}
            H^{\text{sub}} = 
            \begin{pmatrix}
                E_{a_0} & w_0 \e^{-\I \phi_0} \\
                w_0 \e^{\I \phi_0} & E_{b_0}
            \end{pmatrix},
        \end{equation}
        where \(E_{j_0} = \braket{j_0|H|j_0}\), and \(\phi_0\) is the phase of the generically complex off-diagonal element. 
        
        \item Construct the unitary rotation $R_0$ that diagonalizes the $H^{\text{sub}}$ submatrix. Applying it to the $\ket{a_0}$, $\ket{b_0}$ basis states, we obtain,
        \begin{subequations}
        \label{eqn:Jac_rot}
        \begin{align}
             \label{eq:6a}
            \ket{a_0} &\to \ket{a_1} =  \cos\tfrac{\eta_0}{2} \ket{a_0} + \e^{\I \phi_0} \sin\tfrac{\eta_0}{2} \ket{b_0} ,\\
            \label{eq:6b}
            \ket{b_0} &\to \ket{b_1} =  \cos\tfrac{\eta_0}{2} \ket{b_0} - \e^{-\I \phi_0} \sin\tfrac{\eta_0}{2} \ket{a_0}.
        \end{align}
        \end{subequations}
        Above, the rotation angle \(\eta_0\) is defined as
        \begin{equation}\label{eqn:angle}
            \tan \eta_0 = \frac{2 w_0}{E_{a_0} - E_{b_0}}.
        \end{equation}
        The other basis elements are not affected. In the rotated basis,
        \begin{equation}
            \braket{a_1|H|b_1} = 0.
        \end{equation}
        In summary, the basis is updated as: $\ket{j_0} \to \ket{j_1} = R_0 \ket{j_0}$ with 
        \begin{align*}
             \ket{j_1} = \left\{
            \begin{array}{l }
                \ket{j_0},\, \text{if } j \neq a \text{ and } j \neq b, \\[10pt]
                \cos\tfrac{\eta_0}{2} \ket{a_0} + \e^{\I \phi_0} \sin\tfrac{\eta_0}{2} \ket{b_0},   \text{if } j = a, \\[10pt]
                \cos\tfrac{\eta_0}{2} \ket{b_0} - \e^{-\I \phi_0} \sin\tfrac{\eta_0}{2} \ket{a_0},  \text{if } j = b.
            \end{array}
            \right.
        \end{align*}
        \item Go back to step~\ref{itm:algorithm} with the matrix $H_{j_1k_1}$. 
    
        \end{enumerate}
        
        The \emph{Jacobi basis} states after $n$ rotations are
        \begin{equation}
            \ket{j_n} = R_{n-1} \cdots R_0 \ket{j_0}.
        \end{equation}
        They converge to the eigenbasis of $H$ in the limit of infinitely many iterations.
        Indeed, it can be shown that the off-diagonal norm 
        \begin{equation}
            \frac{1}{N} \sum_{j \neq k} |\!\braket{j_n|H|k_n}\!|^2,
            \label{eqn:Gamma_def}
        \end{equation}
        converges to zero exponentially fast in the number of iterations with a rate of at least \(1/N^2\)~\cite{schonhage1964quadratischen,Long2023Beyond}. Thus, the Jacobi algorithm diagonalizes a Hermitian matrix in $\mathcal{O}(N^3)$ floating-point operations. [There are \(\order{N^2}\) rotations, and each involves \(\order{N}\) addition and multiplication operations.]

   The Jacobi algorithm, acting on matrices, is naturally basis-dependent. That is, the rotations $R_n$ depend on the basis in which a fixed operator $H$ is initially represented. In this article, we consider $H = H_0 + JV$, and write $H$ in the eigenbasis of $H_0$. The Jacobi rotations then rotate the eigenbasis of $H_0$ to that of $H$.
        
    \subsection{The Out-of-Equilibrium Eigenstate Thermalization Hypothesis}\label{subsec:ETHspectralforms}
  
    In this subsection, we review the ETH and its extension, known as the out-of-equilibrium ETH, to describe quench dynamics~\cite{foini2024outofequilibriumeigenstatethermalizationhypothesis,Jindal_2024}. We also relate the various form factors in the ETH ansatz to physical quantities.

    Consider the ansatz for the matrix elements of the operator $A$ in Eq.~\eqref{Eq:ETHOffDiag}. The function $A(E)$ in Eq.~\eqref{Eq:ETHOffDiag} is the expectation value of $A$ in the microcanonical ensemble; this follows from direct substitution.

    Next, $|f_A(E,\omega)|^2$ is the spectral function of $A$ in the appropriate microcanonical ensemble. For simplicity, consider infinite temperature.  The Lehmann representation of the infinite temperature autocorrelator $\braket{A(t)A(0)}_{H_0}$ is, 
    \begin{align}\label{eq:Autocor}
        \braket{A(t)A(0)}_{H_0} = \frac{1}{N}\sum_{i_0,j_0} \e^{-\I(E_{i_0}-E_{j_0})t}\braket{j_0|A|i_0}\braket{i_0|A|j_0}.
    \end{align}
    Inserting the ETH ansatz Eq.~\eqref{Eq:ETHOffDiag} into Eq.~\eqref{eq:Autocor},
    the autocorrelator is given by
    \begin{align}
    \begin{split}
      \braket{A(t)A(0)}_{H_0} &=\frac{1}{N}\int \mathrm{d}E\, \nu(E) A^2(w_0,E)\\&+\frac{1}{N}\int \mathrm{d}E \,\nu(E)\int \mathrm{d}\omega\, |f_A(w_0,E,\omega)|^2 \mathrm{e}^{\mathrm{i}\omega t}.
    \end{split}
    \end{align}
    
    The first term on the right-hand side is the diagonal contribution
    to the autocorrelator and determines its infinite-time average. The second term identifies $|f_A(w_0,E,\omega)|^2$ with the Fourier transform of the connected correlator in the microcanonical ensemble at energy density corresponding to infinite temperature, and thus the spectral function. 
    While the various form factors can be defined without any reference to ETH, the latter implies that the matrix elements between individual eigenstates are set by the form factors up to exponentially small corrections in system volume. 
    Given the ETH ansatz, $A(w_0,E)$ and $|f_A(w_0,E,\omega)|^2$ can be extracted as averages over matrix elements:
    	\begin{align}\label{eq:definitions}
        A(w_0,E)\mathrm dE&= \frac{1}{\nu(E)}\sideset{}{'}\sum_{i_0} \braket{i_0|A|i_0}\\
        |f_A(w_0,E,\omega)|^2 \mathrm{d} E \mathrm{d} \omega &=  \frac{1}{{\nu(E) }}\sideset{}{''}\sum_{i_0,j_0} |\braket{i_0|A|j_0}|^2 .
	\end{align}
    
    To lighten the notation in the sums, we have introduced the definitions
    \begin{align}
    \begin{split}
        \sideset{}{'}\sum_i&=\sum_{\substack{i:E_i \in [E, E+\mathrm{d}E)}}\\
        \sideset{}{''}\sum_{i,j}&=\sum_{\substack{i:E_i \in [E, E+\mathrm{d}E)\\j;E_j \in [E+\omega, E+\omega+\mathrm{d}\omega)\\i\neq j}}.
    \end{split}
    \end{align}

    The ETH ansatz in Eq.~\eqref{Eq:ETHOffDiag} needs to be extended to describe transient dynamics~\cite{Pappalardi2022Eigenstate,Jindal_2024}.  
    In the eigenbasis of the Hamiltonian $H_0$, the expectation value of \(A\) is, starting from the initial state $\rho$,
\begin{align}\label{eq:Dynamics}
\begin{split}
		\braket{A(t)}_{H_0} &= \sum_{i_0,j_0} \e^{-\I(E_{i_0}-E_{j_0})t} \braket{i_0|\rho|j_0}\braket{j_0|A|i_0}\\&=\sum_{i_0,j_0} B_{i_0j_0} \e^{-\I(E_{i_0}-E_{j_0})t},
\end{split}
\end{align}
where we introduce the quantity $B_{i_0j_0}\coloneqq\rho_{i_0j_0}A_{j_0i_0}$. 
The extended ETH ansatz is:
\begin{subequations}\label{Eq:Spectralfunctions}
\begin{align}
\rho_{i_0j_0}&=\frac{p(w_0,E)}{\nu(E)}\delta_{i_0j_0}+\frac{g_p(w_0,E,\omega)}{\nu(E)\sqrt{\nu(E+\omega)}}\tilde{R}_{i_0j_0},\\
B_{i_0j_0}&=\frac{B(w_0,E,\omega)}{\nu(E)\nu(E + \omega)} G_{i_0j_0}.
\end{align}
\end{subequations}
Above, $\tilde{R}_{i_0j_0}$ is a pseudo-random number with zero mean and unit variance, $G_{i_0j_0}$ is a pseudo-random number with unit mean, while its variance is not specified by
the ansatz. $B(w_0,E,\omega)$, $p(w_0,E)$ and $g_p(w_0,E,\omega)$ are smooth functions of their (previously defined) arguments. 
The pseudo-random numbers, $\tilde{R}_{i_0j_0}$ in Eq.~\eqref{Eq:Spectralfunctions} and $R_{i_0j_0}$ in Eq.~\eqref{Eq:ETHOffDiag},  are correlated; this is why $B_{i_0j_0}$ has a non-zero mean.

The smooth functions in Eq.~\eqref{Eq:Spectralfunctions} once again encode simple physical quantities. First, $p(E)$ is the probability density of the initial state in the eigenbasis of $H_0$. It determines the mean energy and energy variance of the state $\rho$ through its moments. Next, $|g_p(w_0,E, \omega)|^2/\nu(E)$ is the Fourier transform of the survival probability (the survival probability is defined as the trace overlap of the time evolved state with the initial state $\Tr[\rho(t) \rho(0)]$). Finally, and most importantly for this article, $B(w_0,E,\omega)$ is the Fourier transform of the expectation value of $A(t)$:
\begin{align}\label{eq:timedyndef}
    \braket{A(t)}_{H_0} - \braket{A(0)}_{H_0}=\int \mathrm{d}E\mathrm{d}\omega B(w_0,E,\omega) \left(\e^{\I \omega t}-1\right).
\end{align}

Similar to Eq.~\eqref{eq:definitions}, the expressions in Eq.~\eqref{Eq:Spectralfunctions} define the quantities $p(w_0,E)$, $g_p(w_0,E,\omega)$ and $B(w_0,E,\omega)$ as averages over matrix elements. For instance,
\begin{align}\label{eq:definitions2}
		B(w_0,E,\omega) &=  \frac{1}{{\mathrm{d} E \mathrm{d} \omega }}\sideset{}{''}\sum_{i_0,j_0} \braket{i_0|\rho|j_0} \braket{j_0|A|i_0},
        \\\label{eq:definitipons3}
        p(w_0,E)&= \frac{1}{\mathrm{d} E}\sideset{}{'}\sum_{i_0} \braket{i_0|\rho|i_0}.
\end{align}
\section{Statistical Jacobi approximation~(SJA)}\label{sec:Stat JAcobi}

The pseudo-randomness in the various ETH ansatze in Sec.~\ref{subsec:ETHspectralforms} suggests that a 
statistical description of the rotations proposed by the Jacobi algorithm is sufficient to capture response functions. In this description, the averaged decimated element $w$ plays the role of an inverse flow time, decreasing from a starting value of $w_0$ to zero as the Jacobi algorithm proceeds.

In Sec.~\ref{subsec:Jac_regimes}, we introduce the \emph{distribution of decimated elements}. This is the number density of rotations between different energies as a function of the averaged decimated element $w$. 
In Sec.~\ref{subsec:Evolutionofspectralforms}, assuming that the ETH holds during the Jacobi flow, we introduce a parametrization of the ETH form factors~[cf. Sec.~\ref{subsec:ETHspectralforms}] as a function of $w$.

\subsection{Distribution of decimated elements}
        \label{subsec:Jac_regimes}
         The \emph{distribution of decimated elements} is given by
        \begin{subequations}
        \begin{align}\label{eq:rhodecdef}
        \begin{split}
            \rho_{\mathrm{dec}}(w, E, E') = \sum_{n} &\delta(w - w_n)\left[ \delta(E-E_{a_n})\delta(E'-E_{b_n})\right. \\&+\left. \delta(E-E_{b_n})\delta(E'-E_{a_n}) \right], 
        \end{split}
        \end{align}
        with
        \begin{align}
            \rho_{\mathrm{dec}}(w) &= \int \mathrm{d}E \mathrm{d}E'\, \rho_{\mathrm{dec}}(w,E,E') = 2 \sum_{n} \delta(w - w_n).
        \end{align}
        \label{eqn:rhodec}
        \end{subequations}
        Here, \(\ket{a_n}\) and \(\ket{b_n}\) are the two states involved in the \(n\)-th iteration of the Jacobi algorithm where $w_n$ is the absolute value of the decimated element. 
        By construction, the distribution has the symmetry \(\rho_{\mathrm{dec}}(w,E,E') = \rho_{\mathrm{dec}}(w,E',E)\).

        Another useful quantity is the density of decimated elements per row at a given energy, which is defined as
        \begin{equation}
        \tilde{\rho}(w, E, \Delta) = \frac{\rho_{\mathrm{dec}}(w,E,E-\Delta)}{\nu(E)}.
        \label{eqn:tilderho_def}
        \end{equation}

        The magnitude of the decimated element $w_n$, averaged over several rotations, decreases monotonically with the number of Jacobi rotations, \(n\)~\cite{Long2023Beyond}. This motivates using the average decimated element $w$ to parametrize the progress of the Jacobi algorithm, instead of the index $n$. The Jacobi algorithm thus induces a statistical flow of various quantities as a function of $w$.

        The scaling of the distribution of decimated elements with increasing Hilbert space dimension $N$ allows us to distinguish two different dynamical regimes~\cite{Long2023Beyond}.
        
        In the \emph{sparse} regime, the largest element in each row scales as $\mathcal{O}(1)$. Decimating one element in each row thus reduces the size of the maximum element in each row by an amount $\mathcal{O}(1)$.
        As a consequence, $\rho_{\mathrm{dec}}(w, E, E')$ scales as 
        \begin{align}
            \rho_{\mathrm{dec}}(w, E, E')=\mathcal{O}(N) \qquad \text{(sparse regime)}.
        \end{align}
        The sparse regime occurs in the physics of (prethermal and fully) many-body localized systems and was studied using the SJA in Ref.~\cite{Long2022phenomenology}.
        In this case, the dynamics is characterized by resonances---decimations associated with large rotation angles $\eta$. 

        In contrast, in the \textit{dense regime}, a generic operator $V$ in the eigenbasis of an ETH-satisfying Hamiltonian $H_0$ is represented by a dense matrix (see Eq.~\eqref{Eq:ETHOffDiag}).
        That is, the off-diagonal elements are similar in size and scale as ${N}^{-1/2}$.
        Reducing the total off-diagonal norm of a row by a finite amount $\mathcal{O}(1)$ requires $N$ decimations per row. This results in a reduction of the largest off-diagonal element from $k_1/\sqrt{N}$ to $k_2/\sqrt{N}$. [\(k_1\) and \(k_2\) are \(\order{1}\) with \(N\)]
        Thus
        \begin{align}\label{Eq:DenseRegime}
         \int_{k_1/\sqrt{N}}^{k_2/\sqrt{N}} \rd w\, w^2 \tilde{\rho}(w,E,\Delta) = \order{1} \qquad\text{(dense regime)}.
        \end{align}
        
        Furthermore, as the size of the matrix elements scales as $1/\sqrt{\nu(E)}$, large rotation angles are rarely encountered~\cite{Long2023Beyond}.

Computing response functions of an operator $A$ requires further coarse-grained information from the Jacobi algorithm: the value of the off-diagonal matrix element $A_{b_na_n}$ between the two states involved in the $n$-th iteration.  We thus introduce the joint number density $\rho_{\mathrm{dec},A}(w,E,E',\alpha)$:
\begin{align}\label{eq:rhodecdefjoint}
		\begin{split}
			&\rho_{\mathrm{dec},A}(w, E, E',\alpha) = \sum_{n}  \delta(w - w_{n})\delta(\alpha-\mathrm e^{-i\phi_n}A_{b_n a_n})\\&\left[ \delta(E-E_{a_n})\delta(E'-E_{b_n})\right. +\left. \delta(E-E_{b_n})\delta(E'-E_{a_n}) \right].
		\end{split}
		\end{align}
 Observe that $\alpha$ accounts for the off-diagonal matrix element multiplied by the complex phase $e^{-i\phi_n}$, where $\phi_n$ is the phase associated with the decimated element $H_{a_n b_n}$. This ensures that $\alpha$ does not change when $\ket{a_n}$ or $\ket{b_n}$ are multiplied by overall phases. Only the combination $\alpha$ appears in the derivation of flow equations in Sec.~\ref {sec:flow}.

As before, we define the number density per row,
\begin{equation}
        \tilde{\rho}_A(w, E, \Delta,\alpha)= \frac{\rho_{\mathrm{dec},A}(w,E,E-\Delta,\alpha)}{\nu(E)}.
        \label{eqn:tilderhoa_def}
\end{equation}
       
\subsection{Form factors during the Jacobi flow}\label{subsec:Evolutionofspectralforms}
Consider the matrices $A_{i_0 j_0}$ and $\rho_{i_0 j_0}$ (written in the eigenbasis of $H_0$). After sufficiently many rotations~[$\mathcal{O}(N^2)$, see Eq.~\eqref{Eq:DenseRegime}], the value of $w$ decreases by a small finite amount $\rd w$. Assuming that the ETH holds after these many rotations, the form factors in Sec.~\ref{subsec:ETHspectralforms} remain smooth functions of $E$ and $\omega$, but acquire smooth dependence on $w$. 

After sufficiently many rotations $n= \mathcal{O}(N^2)$, the $w$-dependent form factors are defined as follows:
\begin{align}\label{Eq:Bflow}
       B(w_n,E,\omega) &= \frac{1}{\mathrm{d} E}\frac{1}{\mathrm{d} \omega} \sideset{}{''}\sum_{i,j} \braket{i_n|\rho|j_n} \braket{j_n|A|i_n},\\\label{Eq:fflow}
       |f_A(w_n,E,\omega)|^2 \mathrm{d} E \mathrm{d} \omega&= \frac{1}{\nu(E)} \sideset{}{''}\sum_{i,j} \braket{i_n|A|j_n} \braket{j_n|A|i_n},\\\label{Eq:Aflow}
       A(w_n,E)\rd E&= \frac{1}{\nu(E)}\sideset{}{'}\sum_i \braket{i_n|A|i_n},\\\label{Eq:pflow}
       p(w_n,E)&= \frac{1}{\mathrm{d} E}\sideset{}{'}\sum_i \braket{i_n|\rho|i_n}.
\end{align}
The form factors for $H_0$ are recovered for $w=w_0$, while the form factors for $H=H_0+JV$ are obtained for $w=w_\infty=0$.

We neglect shifts in the energy levels during the Jacobi flow for simplicity.
In principle, they can also be included~\cite[Appendix~\ref{sec:Linear repsonse}]{Long2023Beyond}.
However, the leading effect on the energy levels is an overall shift of the average energy of the initial state, which does not affect dynamics at infinite temperature.
More broadly, the effects of energy level motion can be controllably computed when the entropy density is a slowly varying function of the energy density (so that the ratio of the density of states at energies that differ by the relevant $\omega$ is close to one). 

\section{Form factor flow equations}
\label{sec:flow}
In this section, we obtain flow equations for the form factors in the eigenbasis of $H$ (introduced in \autoref{subsec:Evolutionofspectralforms}) using the SJA.
The initial conditions for the flow equations at $w=w_0$ are given by the form factors for $H_0$. Additionally, the equations explicitly depend on the densities of decimated elements $\rho_{\mathrm{dec}}(w,E,E')$ and $\rho_{\mathrm{dec},A}(w,E,E',\alpha)$. Given these densities, the solution to the flow equation at $w=0$ leads to predictions of the form factors (and thus response functions) with respect to $H=H_0+JV$, as explained in Sec.~\ref{subsec:ETHspectralforms}.

The derivation of the flow equations is provided in Sec.~\ref{subsec:flow_deriv}. In Sec.~\ref{subsec:basic_props}, we discuss a few basic properties of the flow equation and show that the thermodynamic limit is well-defined.

\subsection{Statement of the equations}
\label{Subsec:StatementFlowEq}

Consider an initial state $\rho$, which is stationary with respect to the unperturbed Hamiltonian $H_0$. It is thus diagonal in the starting basis. 

The flow equation for $B(w,E,\omega)$ is given by
\begin{align}\label{Eq:resultshort}
\begin{split}
    -\partial_w B(w,E,\omega)&=F_1[B](w,E,\omega)+F_2[B](w,E,\omega)\\&+G[A,p](w,E,\omega)+D[p](w,E,\omega).
\end{split}
\end{align}
Defining
    \begin{align}\label{eq:Kernel}
    K(w,E,\Delta)= \sin^2\tfrac{\eta(\Delta,w)}{2} \tilde{\rho}(w, E,\Delta),
    \end{align}
the functionals $F_1[B](w,E,\omega)$ and $F_2[B](w,E,\omega)$ have the form
\begin{align}\label{eqn:flowBterms}
        \begin{split}
		&F_1[B](w,E,\omega) = \int \mathrm{d}\Delta \,K(w,E,\Delta)\\&\times\left[\frac{ \nu(E)}{\nu(E-\Delta)}B(w,E-\Delta,\omega+\Delta)-B(w,E,\omega) \right],
        \end{split}
\end{align}
and
\begin{align}
        \begin{split}
        &F_2[B](w,E,\omega) = \int \mathrm{d}\Delta \, K(w,E+\omega,\Delta)\\&\times\left[\frac{\nu(E+\omega)}{\nu(E-\Delta+\omega)}B(w,E,\omega-\Delta)-B(w,E,\omega) \right].
        \end{split}
\end{align}

The term $G[A,p](w,E,\omega)$ is given by
\begin{multline}
		  G[A,p](w,E,\omega)=\\K(w,E,-\omega)
		\left(A(w,E)-A(w,E+\omega)\right)\\
        \times\left(p(w,E)-\frac{\nu(E)}{\nu(E+\omega)}p(w,E+\omega)\right).
\end{multline}
It acts as a source term to the linear integro-differential equation Eq.~\eqref{Eq:resultshort}.

The source term $D[p](w,E,\omega)$ is a consequence of the correlations between the matrix elements of the perturbation $V$ and the observable $A$:
	\begin{multline}\label{eq:Linearresponse}
		D[p](w,E,\omega)=\\ -L(w,E,-\omega)
		\times\left(p(w,E)-\frac{\nu(E)}{\nu(E+\omega)}p(w,E+\omega)\right).
	\end{multline}	
	with 
	\begin{align}
		L(w,E,\omega)=\int \rd \alpha \,\alpha \sin\tfrac{\eta(\omega,w)}{2} \tilde{\rho}_A(w, E,\omega,\alpha).
	\end{align}

The flow equations for  $A(w,E)$ and $p(w,E)$ are given by
\begin{multline}
        -\partial_w p(w,E) = \int \mathrm{d}\Delta K(w,E,\Delta)\\\times\left[p(w,E-\Delta) \frac{\nu(E)}{\nu(E-\Delta)} - p(w,E) \right],
        \label{eqn:flowPhere}
\end{multline}
and 
\begin{multline}
        -\partial_w A(w,E) = \int \mathrm{d}\Delta\, \Big(K(w,E,\Delta)\\\times\left[A(w,E-\Delta)- A(w,E) \right]+2 L(w,E,\Delta)\Big).
        \label{eqn:flowA}
\end{multline}
The flow equation for the spectral function for $|f_A(w,E,\omega)|^2$ defined in Eq.~\eqref{Eq:fflow} is given by (suppressing the arguments $w,E,\omega$)
     \begin{align}\label{Eq:flowAutocor}
     \begin{split}
        -\partial_w |f_A|^2&=F_1\left[|f_A|^2\right]+F_2\left[|f_A|^2\right]+G_A[A]+D_A[A], 
    \end{split}
    \end{align}
with $G_A[A](w,E,\omega)$ defined as
\begin{align}
		  G_A[A](w,E,\omega)=K(w,E,-\omega)
		\left(A(w,E)-A(w,E+\omega)\right)^2.
\end{align}
and
\begin{align}
		  D_A[A](w,E,\omega)=-2 L(w,E,-\omega)
		\left(A(w,E)-A(w,E+\omega) \right).
\end{align}

Note that all the integro-differential equations are linear, with the flow equations for $B(w, E, \omega)$ and $|f_A(w,E,\omega)|^2$ being inhomogeneous.
Any nonlinear dependence on the perturbation $JV$ enters through the decimated density of states $\rho_{\mathrm{dec}}(w,E,E')$ and $\rho_{\mathrm{dec},A}(w,E,E',\alpha)$. A consistent evolution equation of  $\rho_{\mathrm{dec}}(w,E,E')$ as a function of $w$ is beyond the scope of this work.

The flow equations only assume that ETH holds along the flow from $H_0$ to $H$. Their structure is universal, while the specifics of a given problem enter through (i) the initial conditions for \(p\), \(A\) and \(B\), and (ii) the form of $\tilde{\rho}(w,E,\Delta)$ and $\tilde{\rho}_A(w,E,\omega,\alpha)$.
	\begin{figure*}
		\centering
		\includegraphics[width=1\textwidth]{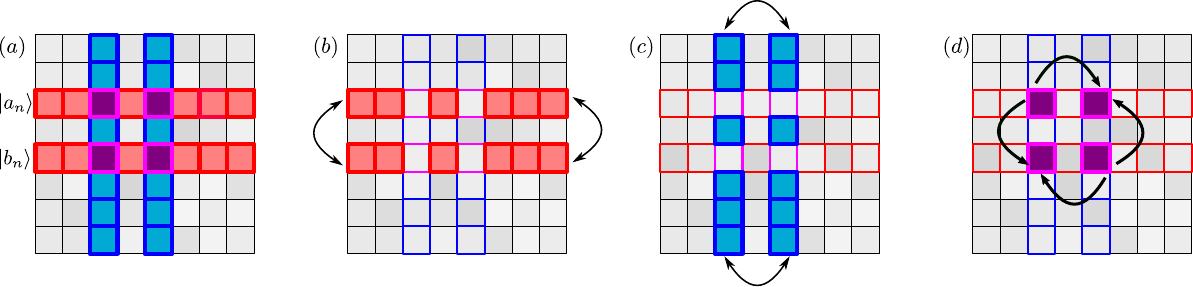}
		\caption{Sketch of one update step for the quantity $B^{(n)}_{ij}=\rho^{(n)}_{ij}A^{(n)}_{ji}$. (a)~The $(n+1)$th update step affects rows~(red) and columns~(blue) with indices $a_n$ and $b_n$. The update can be split into three contributions: (b) Rotations between elements $B^{(n)}_{aj}$ and $B^{(n)}_{bj}$ with $j\neq a,b$ (shown in red). (c) Rotations between elements $B^{(n)}_{ia}$ and $B^{(n)}_{ib}$ with $i\neq a,b$ (shown in blue). Contributions (b) and (c) lead to the terms $F_1[B](w,E,\omega)$, $F_2[B](w,E,\omega)$ in the flow equation Eq.~\eqref{Eq:resultshort}. (d) The third contribution, shown in purple, accounts for the update of the elements $B^{(n)}_{ab}$ and $B^{(n)}_{ba}$, leading to the term $G[A,p](w,E,\omega)$ in Eq.~\eqref{Eq:resultshort}. Additionally, there can be correlations between $A^{n}_{ij}$ and $w_n$. These have to be included separately and lead to the term $D[p](w,E,\omega)$ in Eq.~\eqref{Eq:resultshort}.}
		\label{fig:sketchLDOs}
	\end{figure*}

The functionals in the flow equation for the two form factors, $B(w,E,\omega)$ and $|f_A(w,E,\omega)|^2$, [written in Eq.~\eqref{Eq:resultshort} and Eq.~\eqref{Eq:flowAutocor} respectively] encode the action of the Jacobi rotations on the matrix elements.
In both equations, $F_1[\cdot](w,E,\omega)$ accounts for the row updates shown in Fig.~\ref{fig:sketchLDOs}~(b), while $F_2[\cdot](w,E,\omega)$ accounts for the column updates in Fig.~\ref{fig:sketchLDOs}~(c). The source term, $G[A,p]$ or $G_A[A]$, accounts for the change of the special off-diagonal term that is decimated in each Jacobi rotation, see Fig.~\ref{fig:sketchLDOs}~(d). This term is different between the two equations as $B_{ij}$ is a product of the off-diagonal matrix elements of two different matrices, $\rho_{ij}$ and $A_{ji}$, while $|A_{ij}|^2$ is determined by the off-diagonal matrix element of a single matrix. Indeed, it follows that $G_A[A] = G[A,A]$ (using that $\nu(E)/\nu(E+\omega) \to 1$ at energy densities corresponding to infinite temperature in the thermodynamic limit).

The term $D[p](w,E,\omega)$ takes into account non-trivial correlations between the observable $A$ and the perturbation $V$.

Similarly, the two flow equations for $p(w,E)$ and $A(w,E)$ are identical in structure, as they are derived from the Jacobi updates to the diagonal of a matrix.

\subsection{Derivation of the flow equations}
\label{subsec:flow_deriv}
This subsection provides the derivation of Eq.~\eqref{Eq:resultshort}, starting from updates for a single Jacobi rotation. The derivation of Eq.~\eqref{Eq:flowAutocor} follows completely analogously.
The flow equation for $p(w,E)$ has already been derived in Ref.~\cite{Long2023Beyond}. The flow equation for $A(w,E)$ is a minor modification of the latter.

This sub-section stands alone, and may be skipped by the reader more interested in the solutions of the flow equations. 

\subsubsection{Matrix elements after a single Jacobi rotation}
Consider a single rotation performed by the Jacobi algorithm. If the Jacobi basis is \(\{\ket{j_n}\}\), and the element \(w_n = \max_{j,k} |\!\braket{j_n|H|k_n}\!| = |\!\braket{a_n|H|b_n}\!|\) is to be decimated, then recall that the nontrivial update to the Jacobi basis is~\eqref{eqn:Jac_rot},
	\begin{subequations}
		\begin{align}\tag{\ref{eq:6a}}
			\ket{a_{n+1}} &=  \cos\tfrac{\eta_n}{2} \ket{a_n} + \e^{\I \phi_n}\sin\tfrac{\eta_n}{2} \ket{b_n}, \\
            \tag{\ref{eq:6b}}
			\ket{b_{n+1}} &=   \cos\tfrac{\eta_n}{2} \ket{b_n} -  \e^{-\I \phi_n}\sin\tfrac{\eta_n}{2} \ket{a_n},
		\end{align}
	\end{subequations}
	with \(\tan {\eta_n} = 2w_n/(E_{a_n} - E_{b_n})\). 
	This leads to the following update for the matrix elements:
 \begin{subequations}
	\begin{align}
		A^{(n+1)}_{aj} &=  \cos\tfrac{\eta_n}{2} A^{(n)}_{aj} + \e^{-\I \phi_n}\sin\tfrac{\eta_n}{2} A^{(n)}_{bj}, & j\neq a,b \label{eq:updateopaa} \\
		A^{(n+1)}_{ia} &=  \cos\tfrac{\eta_n}{2} A^{(n)}_{ia} +  \e^{\I \phi_n} \sin\tfrac{\eta_n}{2} A^{(n)}_{ib}, & i\neq a,b \label{eq:updateopab} \\
        \label{eq:updateopac}
		A^{(n+1)}_{ab} &=  \cos^2\tfrac{\eta_n}{2} A^{(n)}_{ab} - \e^{-2\I \phi_n}
        \sin^2\tfrac{\eta_n}{2} A^{(n)}_{ba}\\\nonumber&+\e^{-\I \phi_n}\cos\tfrac{\eta_n}{2}\sin\tfrac{\eta_n}{2} (A^{(n)}_{bb}-A^{(n)}_{aa}), \\
        \label{eq:updateopad}
		A^{(n+1)}_{aa} &=  \cos^2\tfrac{\eta_{n}}{2} A^{(n)}_{aa} + \sin^2\tfrac{\eta_n}{2} A^{(n)}_{bb}\\\nonumber&+\cos\tfrac{\eta_n}{2}\sin\tfrac{\eta_n}{2} (\e^{-\I \phi_n}A^{(n)}_{ba}+\e^{\I \phi_n} A^{(n)}_{ab})\ .
	\end{align}
\end{subequations}
To update $A_{bj}^{(n)}$, exchange the indices $a$ with $b$.
The updates for the matrix elements of $\rho$ are obtained by replacing \(A \to \rho\) in the formulae above.

The quantity $B^{(n)}_{ij} = \rho_{ij}^{(n)} A_{ji}^{(n)}$ is updated if one or both of the indices $i$, $j$ coincide 
with the rotated elements $a$, $b$. We distinguish three different cases. In the first case, only the row index $i$ coincides with one of the two basis elements affected by the Jacobi step, as shown in Fig.~\ref{fig:sketchLDOs}~(b). E.g. for $i=a$, $j\neq a,b$,
\begin{subequations}
	\begin{multline}
		B^{(n+1)}_{aj}-B^{(n)}_{aj} =  \sin^2\tfrac{\eta_n}{2} (B^{(n)}_{bj}-B^{(n)}_{aj}) \\ +\sin\tfrac{\eta_n}{2}\cos\tfrac{\eta_n}{2}(\e^{\I \phi_n}\rho^{(n)}_{aj} A^{(n)}_{jb}+\e^{-\I \phi_n}\rho^{(n)}_{bj} A^{(n)}_{ja}). 
    \label{eq:Ba}
    \end{multline}

In the second case, only the column index $j$ coincides with one of the two basis elements affected by the Jacobi rotation, as shown in Fig.~\ref{fig:sketchLDOs}~(c). E.g., for $i\neq a,b$, $j=a$:
\begin{multline}
		B^{(n+1)}_{ia}-B^{(n)}_{ia} =  \sin^2\tfrac{\eta_n}{2} (B^{(n)}_{ib}-B^{(n)}_{ia})\\
        +\sin\tfrac{\eta_n}{2}\cos\tfrac{\eta_n}{2}(\e^{-\I \phi_n}\rho^{(n)}_{ia} A^{(n)}_{bi}+\e^{\I \phi_n}\rho^{(n)}_{ib} A^{(n)}_{ai}).\label{eq:Bb}
\end{multline}

In the third case, both indices $i$ and $j$ coincide with the basis elements affected by the Jacobi rotation (with $i \neq j$). E.g., for $i=a$, $j=b$: 
\begin{widetext}
\begin{multline}\label{eq:flowequationstart2}
	B^{(n+1)}_{ab}-B^{(n)}_{ab}=	 \cos^2\tfrac{\eta_{n}}{2}\sin^2\tfrac{\eta_{n}}{2}\left(-2B^{(n)}_{ab}-\e^{2 \I \phi_n}\rho^{(n)}_{ab}A^{(n)}_{ab}-\e^{-2 \I \phi_n}\rho^{(n)}_{ba}A^{(n)}_{ba}+(\rho^{(n)}_{aa}-\rho^{(n)}_{bb})(A^{(n)}_{aa}-A^{(n)}_{bb})\right)
  \\+\sin^4\tfrac{\eta_n}{2}(B^{(n)}_{ba}-B^{(n)}_{ab})+\cos^3\tfrac{\eta_{n}}{2}\sin\tfrac{\eta_{n}}{2}\big(\e^{-\I \phi_n}(\rho^{(n)}_{bb}-\rho^{(n)}_{aa}) A^{(n)}_{ba}+\e^{\I \phi_n}\rho^{(n)}_{ab}(A^{(n)}_{bb}-A^{(n)}_{aa})\big)
  	\\+\cos\tfrac{\eta_n}{2}\sin^3\tfrac{\eta_n}{2}\big(\e^{\I \phi_n}(\rho^{(n)}_{aa}-\rho^{(n)}_{bb}) A^{(n)}_{ab}+\e^{-\I \phi_n}\rho^{(n)}_{ba}(A^{(n)}_{aa}-A^{(n)}_{bb})\big).
\end{multline}
\end{widetext}

\end{subequations}

\subsubsection{Averaging over multiple rotations}\label{subsubsec:important}

Recall that we work in the \emph{dense} regime of the Jacobi algorithm, as discussed in Sec.~\ref{subsec:Jac_regimes}, and thus the rotation angle $\eta_n$ is small [$\eta_n = \mathcal{O}(1/\sqrt{N})$] and of random sign. This justifies dropping the third- and fourth-order contributions in $\sin \tfrac{\eta_n}{2}$. To be consistent within this approximation, we also set $\cos^2\tfrac{\eta_n}{2}\sin^2\tfrac{\eta_n}{2}\approx \sin^2\tfrac{\eta_n}{2}$.

Second, as $B^{(0)}_{i j}$ is initially diagonal (because \(\rho\) is diagonal), $B^{(n)}_{ab}$ is suppressed by a factor $\sin^2 \eta = \mathcal{O}(1/N)$ in comparison to the diagonal elements $\rho^{(n)}_{ii}A^{(n)}_{ii}$. 
This justifies dropping the term $-2B^{(n)}_{ab}-\e^{2 \I \phi_n}\rho^{(n)}_{ab}A^{(n)}_{ab}-\e^{-2 \I \phi_n}\rho^{(n)}_{ba}A^{(n)}_{ba}$ in the first line of Eq.~\eqref{eq:flowequationstart2}.

We assume that subsequent Jacobi iterations are uncorrelated.
If \(A\) is initially diagonal, then off-diagonal terms \(A_{ba}^{(n)}\) can only be generated by these uncorrelated rotations. Thus, the terms in the update equations involving the combination
\[
\e^{-\I \phi_n}\,\sin\!\left(\tfrac{\eta_n}{2}\right) A^{(n)}_{ba}
\]
typically have no definite sign and zero mean, as both \(\sin(\eta_n/2)\) and \(\e^{-\I \phi_n} A^{(n)}_{ba}\) are assumed uncorrelated, and \(\sin(\eta_n/2)\) can be either positive or negative. As such, the sum of these terms is suppressed by a factor
of $\tfrac{1}{\sqrt{N}}$ in comparison to quadratic terms $\sin^2 (\eta/2)$, which always come with a phase determined by \(\e^{-\I \phi_n}A^{(n)}_{ba}\).
We assume that the terms linear in \(\sin(\eta/2)\) can be neglected when \(A\) is initially diagonal (and our numerics support this assumption, Sec.~\ref{sec:numerics}).
In contrast, if $A$ initially contains off-diagonal elements, the linear terms encode correlations between $A$ and the perturbation $V$.  
As discussed in Sec.~\ref{sec:flow_sol}, their inclusion is essential to reproduce linear response theory and corrections thereof.

Taking all these approximations into account, the update of $B^{(n)}_{ij}$ for $i \neq j$ under $\mathrm{d}n$ consecutive Jacobi iterations can be expressed as~(noting that the indices $a=a_m$, $b=b_m$ depend implicitly on the Jacobi step)
\begin{widetext}
\begin{subequations}\label{eq:Bequationdiscrete}
\begin{align}\label{eq:Terma}
    B^{(n+\rd n)}_{ij}-B^{(n)}_{ij}\approx\sum_{n\leq m< n+\mathrm{d}n} \sin^2\tfrac{\eta_m}{2}&[(B^{(m)}_{aj}-B^{(m)}_{ij})\delta_{i, b}(1-\delta_{j,a})(1-\delta_{j,b})+(a\leftrightarrow b)\\+&(B^{(m)}_{ia}-B^{(m)}_{ij}) \delta_{j, b}(1-\delta_{i,a})(1-\delta_{i,b}) +(a\leftrightarrow b)\\+&(\rho^{(m)}_{ii}-\rho^{(m)}_{jj})(A^{(m)}_{ii}-A^{(m)}_{jj})\left(\delta_{i,a}\delta_{j,b}+(a\leftrightarrow b)\right)]\\\label{eq:cor1}+&\e^{-\I \phi_m}\sin\tfrac{\eta_{m}}{2}(\rho^{(m)}_{bb} A^{(m)}_{ba}-\rho^{(m)}_{aa} A^{(m)}_{ba}).
\end{align}
\end{subequations}
\end{widetext}
Each line in Eq.~\eqref{eq:Bequationdiscrete} corresponds to the three cases Eq.~\eqref{eq:Ba}, Eq.~\eqref{eq:Bb} and Eq.~\eqref{eq:flowequationstart2}, respectively. Additionally, Eq.~\eqref{eq:cor1} contains terms linear in $\sin\tfrac{\eta_{m}}{2}$, which capture correlations between $A$ and $V$ in Eq.~\eqref{eq:flowequationstart2}.

\subsubsection{Parametrization by the decimated element}
\label{subsec:dec_ele_flow}
    
We assume that a continuous, monotonically decreasing parameter $w$ can replace $w_n$. As shown in~\cite{Long2023Beyond}, this is possible if we average over multiple rotations and, in addition, assume that $B^{(n)}_{ij}$, $\rho^{(n)}_{ij}$ and $A^{(n)}_{ij}$ vary slowly with $w_n$. 
Supposing that \(\mathrm{d}n\) rotations reduce the decimated element from \(w\) to \(w - \mathrm{d}w\), the left hand side of Eq.~\eqref{eq:Bequationdiscrete} becomes 
	\begin{equation}
		B^{(n+\mathrm{d}n)}_{ij}-B^{(n)}_{ij} \to -\partial_w B_{ij}(w) \mathrm{d} w,
		\label{eqn:w_replace}
	\end{equation}
	where we take \(\mathrm{d} w\) to be infinitesimal, and $B_{ij}(w_n) = B^{(n)}_{ij}$.
	
	\subsubsection{Using the distribution of decimated elements}
	\label{subsec:integral_replace}
	We now replace the sum in Eq.~\eqref{eq:Bequationdiscrete} with an integral.
    The number of rotations performed between states of energy $E_{a_m} \in [E, E+\mathrm{d}E)$ and $E_{b_m} \in [E-\Delta, E-\Delta+\mathrm{d}\Delta)$ while $w_m \in (w-\mathrm{d}w,w]$, is given by the distribution of decimated elements
	\begin{equation}
		\rho_{\mathrm{dec}}(w, E, E-\Delta)\, \mathrm{d} w \,\mathrm{d}E \, \mathrm{d}\Delta \; .
		\label{eqn:num_dens_rot}
	\end{equation}
	To obtain the average number of rotations for a single state $\ket{a_{i}}$ in the interval $E_{i}\in[E, E+\mathrm{d}E)$, we divide Eq.~\eqref{eqn:num_dens_rot} by the number of states in this shell, $\nu(E) \mathrm{d}E$, which gives $\tilde{\rho} \,\mathrm{d}w\mathrm{d}\Delta$ as in Eq.~\eqref{eqn:tilderho_def}. 
    After replacing the sum in Eq.~\eqref{eq:Bequationdiscrete} by an integral, this gives
	\begin{equation}
		\sum_m \sin^2\tfrac{\eta_m}{2}[\dots] \to \mathrm{d} w \int \sin^2\tfrac{\eta(\Delta,w)}{2} \mathrm{d}\Delta\, \tilde{\rho}(w, E,\Delta)[\dots].
		\label{eqn:sum_rep}
	\end{equation}
    and 
    \begin{multline}
		\sum_m \e^{-\I \phi_m}\sin\tfrac{\eta_m}{2}A_{ba}[\dots] \\\to \mathrm{d} w \int \sin \tfrac{\eta(\Delta,w)}{2} \mathrm{d}\Delta\, \mathrm{d}\alpha\, \alpha \tilde{\rho}_A(w, E,\Delta,\alpha)[\dots].
		\label{eqn:sum_rep}
	\end{multline}
	As an illustration, the term on the right-hand side of Eq.~\eqref{eq:Terma} transforms to: 
\begin{align}\label{eq:Bequationdiscretestep1}
\begin{split}
    \text{Eq.~\eqref{eq:Terma}}\rightarrow \mathrm{d} w 
&\int \sum_{k\neq j} \mathrm{d}\Delta\, K(w,E_i,\Delta)\\& (B_{kj}(w)-B_{ij}(w))\delta(E_i-\Delta-E_k),
\end{split}
\end{align}
where in the last line we use the definition of the kernel $K(w,E,\Delta)$ in Eq.~\eqref{eq:Kernel}.

\subsubsection{Reintroducing form factors}
\label{subsec:prob_to_dens}

Finally, we replace the terms $\rho_{ii}(w)$, $A_{ii}(w)$ and $B_{ij}(w)$ with averages over small energy windows, using Eqs.~(\ref{Eq:Bflow}-\ref{Eq:pflow}).

   For illustration, the term Eq.~\eqref{eq:Bequationdiscretestep1} transforms as 
  \begin{multline}\label{eq:Bequationdiscretestep2}
 \text{Eq.~\eqref{eq:Bequationdiscretestep1}}
 \rightarrow \,\mathrm{d} w \int \mathrm{d}\Delta\, K(w,E_i,\Delta)\\\left(\frac{B(w,E_i-\Delta,\omega+\Delta)}{\nu(E_i-\Delta)\nu(E_i+\omega)}-\frac{B(w,E_i,\omega)}{\nu(E_i)\nu(E_i+\omega)}\right)\\=
 \frac{\mathrm{d} w}{\nu(E_i)\nu(E_i+\omega)}F_1[B](w,E_i,\omega)\; .
\end{multline}
   
	Applying the same substitutions to the other terms, Eq.~\eqref{eq:Bequationdiscrete} transforms to a flow equation for $B(w,E,\omega)$.
 \begin{align}\label{eqn:flowBend}
 \begin{split}
     -\frac{\partial_w B(w,E,\omega)}{\nu(E)\nu(E+\omega)}&=\frac{1}{\nu(E)\nu(E+\omega)}\Bigl(F_1[B](w,E,\omega)\\&+F_2[B](w,E,\omega)+G[A,p](w,E,\omega)\\&+
     D[p](w,E,\omega)\Bigr).
 \end{split}
 \end{align}
Multiplying both sides by $\nu(E)\nu(E+\omega)$, we recover Eq.~\eqref{Eq:resultshort}. 

With the same assumptions, the other flow equations in Sec.~\ref{Subsec:StatementFlowEq} can be derived. The derivation neglects possible shifts of the energy levels during the flow equation. See the discussion in Sec.~\ref{subsec:Evolutionofspectralforms}. 

	\subsection{Basic properties of the flow equations}
	\label{subsec:basic_props}
    The flow equations in Eq.~\eqref{Eq:resultshort} have trivial fixed points and a well-defined thermodynamic limit.
	
	\subsubsection{Trivial fixed points}
	When the operator $A$ is the identity, it has no dynamics. Consequently, we expect $f_A(w, E, \omega) = 0$, $B(w, E, \omega)=0$, and $A(w,E) = \textrm{const}$. This is indeed a solution to the flow equations in Sec.~\ref{Subsec:StatementFlowEq} for any choice of $p(w=w_0, E)$. 

    Similarly, when the initial density matrix is the identity, it is unaffected by Jacobi rotations and $\langle A(t) \rangle = \textrm{const}$ for all $t$. This is reflected in $p(w,E)=\nu(E)/N$ being a solution of Eq.~\eqref{eqn:flowPhere} and $B(w, E, \omega)=0$ being a solution to Eq.~\eqref{Eq:resultshort}. Note that $|f_A|^2(w, E, \omega)$ can still be non-trivial in this case because of the source term in Eq.~\eqref{Eq:flowAutocor}. 
    
	\subsubsection{Thermodynamic limit}\label{sec:ThermodynamicLimit}
    
    The thermodynamic limit is defined as the limit of $N\to \infty$. We expect that the form factors (that determine response functions) are finite in this limit. Here, we establish that the flow equations in Sec.~\ref{Subsec:StatementFlowEq} have a finite thermodynamic limit. We present the argument for the terms that are quadratic in $\sin \frac{\eta}{2}$; the corresponding argument for the term $D[p](w,E,\omega)$ is analogous.    
    
    First, the ratio of the density of states $\nu(E)/\nu(E+\omega)$ is finite as $N\to \infty$. Thus, all we need to show is that the function $K(w, E, \Delta)$ has a well-defined limit.

    The function $K(w, E, \Delta)$ is a product of two terms. Consider the first, $\sin^2(\eta(\Delta,w)/2)$. In the dense regime, the decimated element $w$ is of order $1/\sqrt{\nu(E)}$, while the typical energy difference $\Delta$ is of order 1. Thus, the rotation angles $\eta(\Delta,w)$ are small, and the small angle approximation is controlled,
    \begin{align*}
        \sin^2 \frac{\eta(\Delta,w)}{2}&=\frac{w^2}{\Delta^2}+\mathcal{O}\left(\frac{w^4}{\Delta^4}\right)
    \end{align*}

    The dependence of the flow equation on the average decimated element $w$ appears then in the combination
    \begin{align}
        w^2\tilde{\rho}(w,E,\Delta).   
    \end{align}
    As discussed in Sec.~\ref{subsec:Jac_regimes}, the integral over this quantity is finite in the dense regime
     \begin{align}\label{Eq:DenseRegime2}
         \int_{k_1/\sqrt{N}}^{k_2/\sqrt{N}} \rd w\, w^2 \tilde{\rho}(w,E,\Delta) = \mathcal{O}(1)
        \end{align}
    and the solutions of the flow equation have a good \(N \to \infty\) limit.
    
    The input from the Jacobi algorithm with a well-defined thermodynamic limit is thus the function $K(w, E, \Delta)$. Should this be provided, the SJA computes response functions in the thermodynamic limit. In well-thermalizing systems, relatively small system sizes are sufficient to compute $K(w, E, \Delta)$ accurately, and obtain SJA solutions in the thermodynamic limit. We discuss finite-size effects in the SJA solutions further in Sec.~\ref{sec:Finitesizescaling}.

	\section{Iterative solution of the flow equation}
	\label{sec:flow_sol}
    The flow equation Eq.~\eqref{Eq:resultshort} is an inhomogeneous, linear integro-differential equation, and difficult to solve for generic initial conditions. 
    In this section, we obtain an iterative solution for a weak perturbation. The leading order of the iterative solution recovers the result of second-order time-dependent perturbation theory. 
    
    To be more specific, consider an expansion in the parameter $\epsilon=J/\sigma_\omega$, with $\sigma_\omega$ denoting the width of $|f_V|^2$ in $\omega$ at infinite temperature.
    Here $|f_V|^2$ denotes the spectral function of the perturbation $V$ in the eigenbasis of $H_0$:
    \begin{align}\label{eq:deffv}
    |f_V(E,\omega)|^2 \mathrm{d} E \mathrm{d} \omega &=  \frac{1}{{\nu(E) }} \sideset{}{''}\sum_{i_0,j_0} |\braket{i_0|V|j_0}|^2.
    \end{align}
    To make the dependence of the flow equation Eq.~\eqref{Eq:resultshort} on $\epsilon$ explicit, we rescale the kernel $K(w,E,\Delta)$ in dimensionless units. 
    Recall that $K(w,E,\Delta) \approx (w^2/\Delta^2)\tilde{\rho}(w, E,\Delta)$.
    The characteristic scale of $\tilde{\rho}(w, E_0,\Delta)$ in the last variable $\Delta$ is specified by the energy difference of states affected by the perturbation $V$ and is thus set by the spectral bandwidth of the perturbation $\sigma_\omega$~\cite{Long2023Beyond}.
    Furthermore, the size of the largest decimated element $w_0$ is characterized by the scale of the perturbation $J$.
    This motivates the introduction of rescaled coordinates $\Delta=\sigma_\omega \xi$, $w=J x$, the definition of $\rho'$, the rescaled decimated number density,
    \begin{align}
        \tilde{\rho}(w, E,\Delta)\mathrm{d}\Delta\mathrm{d}w= \rho'(x,E,\xi)\mathrm{d}\xi\mathrm{d}x 
    \end{align}
    Together with $w^2/\Delta^2=\epsilon^2 x^2/\xi^2$, this leads to the rescaled kernel 
    \begin{align}
        K(w,E,\Delta) \mathrm{d}\Delta\mathrm{d}w=\epsilon^2 K'(x,E,\xi)\mathrm{d}\xi\mathrm{d}x 
    \end{align}
    In the rescaled variables, the flow equation has the form
     \begin{multline}\label{eqn:flowBshorter}
     -\partial_x B(x,E,\omega)=\epsilon^2 \left\{ F'_1[B](x,E,\omega)+F'_2[B](x,E,\omega)\right.\\
     \left.+G'[A,p](x,E,\omega)\right\}+\epsilon D'[p](x,E,\omega).
     \end{multline}
As an explicit example, $F'_1[B](x,E,\omega)$ is given by 
\begin{align}\label{eqn:flowBterms}
        \begin{split}
		&F'_1[B](x,E,\omega) = \int \mathrm{d}\xi \,K'(x,E,\xi)\\&\times\left[\frac{ \nu(E)}{\nu(E-\xi \sigma_\omega)}B(x,E-\sigma_\omega\xi,\omega+\sigma_\omega\xi)-B(x,E,\omega) \right],
        \end{split}
\end{align}
and the other terms in the flow equations Eq.~\eqref{Eq:resultshort}, Eq.~\eqref{eqn:flowPhere}, and Eq.~\eqref{eqn:flowA} are rescaled similarly.

We distinguish two cases. First, we consider the case where $A$ is initially diagonal, i.e., $D'[p](x_0,E,\omega)=0$. As discussed in Sec.~\ref {subsubsec:important}, the term for $D'[p](x,E,\omega)$ is suppressed during the entire Jacobi algorithm by a factor $\frac{1}{\sqrt{N}}$, and it is justified to neglect it in this case.
The right-hand side of the differential equation is suppressed by a factor $\epsilon^2$.
The set of equations can thus be solved iteratively order by order in $\epsilon^2$.
To do so, consider sequences $p_k(x,E)$, $A_k(x,E)$ and $B_k(x,E,\omega)$ for $k\geq 0$ and initial conditions at $k=0$, $p_{0}(x,E)=p(x_0,E)$ etc. We define,
\begin{multline}\label{eq:iterative}
-\partial_x B_{k+1}(x,E,\omega)=\epsilon^2 \Bigl( F'_1[B_k](x,E,\omega)\\+F'_2[B_k](x,E,\omega)+G'[A_k,p_k](x,E,\omega)\Bigr),
\end{multline}
and similarly for $p_k(x,E)$ and $A_k(x,E)$.
Furthermore, the first elements of these sequences are given by
    \begin{align}
        p_{0}(x,E)&=p(x_0,E) \\
        A_{0}(x,E)&=A(x_0,E)\\
        B_{0}(x,E,\omega)&=B(x_0,E,\omega)=0.
    \end{align}
In the following, we explicitly solve the first order correction in $\epsilon^2$.
Since $B_{0}(x,E,\omega)=0$, $F'_1[B_0](x,E,\omega)=F'_2[B_0](x,E,\omega)=0$, the flow equation at lowest order simplifies to
\begin{align}
-\partial_x B_{1}(x,E,\omega)=\epsilon^2G'[A_{0},p_{0}](x,E,\omega).
\end{align}
Integrating and re-instating unscaled variables, 
\begin{align}\label{eq:solution}
\begin{split}
	&B_1(w=0,E,\omega)=J^2\frac{|f_{\mathrm{Jac}}(E,\omega)|^2}{\omega^2}	\\& \Biggl(\Bigl(p(w_0,E)-\frac{\nu(E)}{\nu(E+\omega)} p(w_0,E+\omega) \Bigr)\\&\times\Bigl(A(w_0,E)-A(w_0,E+\omega)\Bigr) \Biggr),
\end{split}
\end{align}
where the Jacobi spectral function $|f_{\mathrm{Jac}}|^2$ is defined as,
\begin{align*}
    J^2 |f_{\mathrm{Jac}}(E,\omega)|^2= \int_0^\infty  \mathrm{d}w\, w^2 \tilde{\rho}(w,E,-\omega).
\end{align*} 

There is an important connection with time-dependent perturbation theory.
The Jacobi spectral function agrees at leading order with the spectral function $|f_{V}(E,\omega)|^2$ for the perturbation $V$ in the basis of $H_0$~\cite{Long2023Beyond}:
\begin{align}\label{eq:fJac}
	|f_{\mathrm{Jac}}(E,\omega)|^2=|f_{V}(E,\omega)|^2+\mathcal{O}\left(\frac{J^2}{\sigma^3_\omega}\right).
\end{align}
Replacing $f_{\mathrm{Jac}}(E,\omega)$ by $f_{V}(E,\omega)$ in Eq.~\eqref{eq:solution}, we recover the result of second-order time-dependent perturbation theory (see App.~\ref{app:Second order perturbation theory}). While the expressions for the first-order solution of the flow equation and second-order time-dependent perturbation theory look qualitatively similar, the replacement of $f_{V}(E,\omega)$ by $f_{\mathrm{Jac}}(E,\omega)$ can already account for large corrections~\cite{Long2023Beyond}.

We now consider the second case, in which \(A\) is not initially diagonal and $D'[p](x,E,\omega)\neq0$. As before, we can iteratively solve the flow equation. However, the leading order is proportional to the dimensionless perturbation parameter $\epsilon$.
Collecting terms with the same scaling with \(\epsilon\),
the iterative equations in this case are
\begin{multline}\label{eq:iterative2}
-\partial_x B_{k+1}(x,E,\omega)=\epsilon D'[p_k](x,E,\omega)+\\\epsilon^2 \Bigl( F'_1[B_{k-1}](x,E,\omega)\\+F'_2[B_{k-1}](x,E,\omega)+G'[A_{k-1},p_{k-1}](x,E,\omega)\Bigr),
\end{multline}

At the lowest order, we obtain 
\begin{align}
-\partial_x B_{1}(x,E,\omega)=\epsilon D'[p_0](x,E,\omega).
\end{align}
Integrating and using un-rescaled variables, this gives
\begin{multline}\label{eq:LinearresponseJac}
    B_1(0,E,\omega)=J \frac{g_{\mathrm{Jac}}(E,\omega)}{\omega}\\\Bigl(p(w_0,E)-\frac{\nu(E)}{\nu(E+\omega)} p(w_0,E+\omega) \Bigr)
\end{multline}
with 
\begin{align}
    J g_{\mathrm{Jac}}(E,\omega)=\int_0^\infty \rd w w \int \rd \alpha\, \alpha\tilde{\rho}_A(w,E,-\omega,\alpha).
\end{align}
To interpret this term, assume that each matrix element is decimated once and not affected by other rotations. Then we obtain
\begin{multline}\label{eq:LRcomparison}
J\frac{g_{\rm Jac}(E,\omega)}{\omega}
\left[p(w_0,E)-\frac{\nu(E)}{\nu(E+\omega)} p(w_0,E+\omega)\right]
\\\approx
J\sum_{a_0,b_0}
\delta\left(\omega-(E_{b_0}-E_{a_0})\right)
\delta(E-E_{a_0})
\\\times
\frac{[\rho_{a_0a_0}V_{a_0b_0}A_{b_0a_0}-\rho_{b_0b_0}V_{a_0b_0}A_{b_0a_0}]
}{E_{b_0}-E_{a_0}}.
\end{multline}
In this case, we have recovered the contribution of linear response~[cf.~App.~\ref{sec:Linear repsonse}]. As before, subsequent rotations can lead to significant deviation between the linear response term and the expression appearing in the first order of the iterative solution in Eq.~\eqref{eq:LinearresponseJac}.  

Finally, we note that the iterative solution is technically a Picard iteration. This method has a guaranteed radius of convergence in the integration variable $w$. We note that this does not directly imply convergence in $t$. Consequently, convergence of the iterative procedure in time is not ensured.

	\section{Numerical tests}
	\label{sec:numerics}
In this section, we compare the results of numerically exact simulations to (i) the iterative solutions of the flow equations [Eq.~\eqref{eq:iterative} and Eq.~\eqref{eq:iterative2}], and (ii) low-order time-dependent perturbation theory~(TDPT). The solutions of the flow equations reproduce features of the exact dynamics at short time scales---where TDPT also performs well---while capturing long-time steady-state values---where TDPT fails. 

\subsection{Numerical implementation}

In this subsection, we describe the observables $A$ in different models, and describe the numerical scheme to solve Eq.~\eqref{eq:iterative} and Eq.~\eqref{eq:iterative2}.
We study the dynamics under a quench experiment.
We initialize the system in a stationary state $\rho$ of the unperturbed Hamiltonian $H_0$ and follow the dynamics of an observable $A$ under the quenched Hamiltonian $H=H_0+J V$. 
We present results for random matrix models and one-dimensional spin-1/2 models. 
By absorbing diagonal elements in $H_0$, $V$ can be made purely off-diagonal.

\begin{figure*}[t!]
		\centering
		\includegraphics[width=1\textwidth]{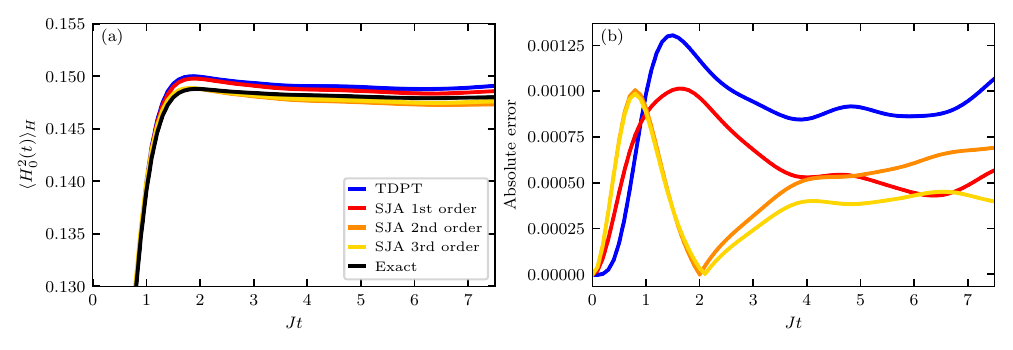}
		\caption{Evolution of $\braket{H_0^2}_H$ under a quench Eq.~\eqref{eq:Quench1}, $J=0.5$, $\sigma_\omega=1.5$, $\epsilon=1/3$, $N=2048$. (a) Comparison of exact time evolution (black), with second-order time-dependent perturbation theory (blue), and the statistical Jacobi approximation to different orders (red, orange, and yellow). The data are averaged over 10 random realizations. The agreement with the exact curve improves at second order. (b) Absolute error $|\braket{H_0^2}_{H,\text{approx}}-\braket{H_0^2}_{H,\text{exact}}|$. The iterative solution outperforms second-order time-dependent perturbation theory at intermediate and long times. } 
		\label{fig:sketchRmatrix}
\end{figure*}

For the random matrix model and the one-dimensional spin-1/2 model, we analyze the dynamics of different observables $A$.
For the random matrix models analyzed in Sec.~\ref{subsec:randommatrix}, we consider $A = H_0^2$, which measures the variance of the energy with respect to the initial Hamiltonian in the time-evolved state (the initial state has \(\expval{H_0} = 0\).) Since we probe dynamics in the middle of the spectrum, $H_0^2$ exhibits more structure in its evolution compared to $H_0$ itself. This observable is diagonal in the eigenbasis of the original Hamiltonian.
In this case, we obtain the iterative solutions using Eq.~\eqref{eq:iterative}.

 For the spin-1/2 model presented in Sec.~\ref{subsec:spinmodel}, we consider the observable  
\begin{align}\label{eq:Z}
    Z=\sum_i \sigma^z_i,
\end{align}
which measures the total magnetization of the system. We choose \(H_0\) that does not commute with \(Z\), so we have to take correlations between the perturbation $V$ and the observable $Z$ into account.
In this case, we have to iteratively solve Eq.~\eqref{eq:iterative2}. 

In order to predict $\braket{A(t)}_H$ from $\braket{A(t)}_{H_0}$, we implement the iterative algorithm described in Sec.~\ref{sec:flow_sol}. 
We determine the initial form factors numerically.
The algorithm produces a sequence of solutions $B_{k}(w_{\text{min}},E,\omega)$ with a cutoff scale $w_{\text{min}}$ defined below, which determines the $k$-th SJA approximation to $\braket{A(t)}_H$:
\begin{align}\label{eq:timedyndef}
    \braket{A(t)}_{H} - \braket{A(0)}_{H}=\int \mathrm{d}E\mathrm{d}\omega B_k(w_{min},E,\omega) \left(\e^{\I \omega t}-1\right).
\end{align}

We run the exact Jacobi algorithm numerically to determine the decimated element $w_n$.
The iterative solutions $B_{k+1}(w_{n+1},E,\omega)$ are updated after each Jacobi iteration as
\begin{align}\label{eq:numerically}
\begin{split}
    &B_{k+1}(w_{n+1},E,\omega)-B_{k+1}(w_n,E,\omega)\\&=F_1[B_k](w_n,E,\omega)+F_2[B_k](w_n,E,\omega)\\&+G[A_k,p_k](w_n,E,\omega)+D[p_k](w_n,E,\omega).
\end{split}
\end{align}
 $\rho_{\mathrm{dec}}(w,E,E')$ and $\rho_{\mathrm{dec},Z}(w,E,E',\alpha)$ are defined as a sum of discrete delta functions as in Eq.~\eqref{eq:rhodecdef} and Eq.~\eqref{eq:rhodecdefjoint}, respectively.
Note that the discrete update equation in Eq.~\eqref{eq:numerically} does not explicitly require the computation of the distribution of decimated elements.
By construction, the zeroth order solution is unchanged by the Jacobi flow, so that $B_{0}(w_{n+1},E,\omega)=B_{0}(w_{n},E,\omega)$.

We coarse-grain the matrix elements over small energy windows: we average all relevant quantities, $B(w_n,E,\omega)$, $A(w_n,E)$, and $\rho(w_n,E)$ over $N_{\mathrm{bin}}=4$ consecutive eigenstates. 
We empirically find that $N_{\mathrm{bin}}=4$ is sufficient to obtain coarse-grained form factors. 
The energy $E$ for each bin is given by the average energy of the four states.
To obtain the initial condition $B_{k}(w_{0},E,\omega)$,  order the states $\ket{i_0}$ by energy, and consider the $m$-th energy bin $E_m$, and the $n$-th $\omega_n$. $B_{k}(w_{0},E_m,\omega_n)$ is then given by 
\begin{align}
\begin{split}
    B_0(w_0,E_m,\omega_n)=&\frac{1}{N_{bin}}\sum_{i=N_{\mathrm{bin}} m+1}^{N_{\mathrm{bin}} (m+1)}
    \sum_{j=N_{\mathrm{bin}} (m+n)+1}^{N_{\mathrm{bin}} ((m+n)+1)}\\& \times \braket{i_0|\rho|j_0} \braket{j_0|A|i_0}
\end{split}
\end{align}
and  $B_k(w_0,E_m,\omega_n)=B_0(w_0,E_m,\omega_n)$. The other form factors $p_k(w_0,E_m)$ and $A_k(w_0,E_m)$ are initialized analogously.

We repeat the update Eq.~\eqref{eq:numerically}, as long as $w_n>w_{\text{min}}$,
where $w_{\text{min}}$ serves as a numerical cutoff scale. In our numerical simulations, we find that the matrix becomes effectively diagonal for $w_{\text{min}} = 10^{-6}$.

\subsection{Random matrix models}\label{subsec:randommatrix}
As a first test, we benchmark our results against two different random matrix models.
These random matrix models feature no correlations between matrix elements in the eigenbasis of $H_0$, justifying many assumptions of the SJA. 

In both models, $H_0$ is diagonal, with elements normally distributed with width 2.5 and mean 0.
The perturbation is purely off-diagonal with matrix elements 
\begin{align}
   V_{jk}=\frac{f_{V}(E,\omega)}{\sqrt{\nu(E+\omega)}} R_{jk}.
\end{align}
Here $R_{jk}$ are independent normal random variables with mean zero and unit variance for $j<k$, \(R_{jk} = R_{kj}\), and $|f_{V}(E,\omega)|^2$ is the spectral function of the perturbation with respect to $H_0$. The chosen form of $f_{V}(E,\omega)$ differs between the two models considered.
We choose the initial state
\begin{align}
    \rho=\frac{1}{\mathcal{N}}\sum_{|E_{i_0}|<0.5}\ketbra{i_0}{i_0},
\end{align}
with $\mathcal{N}$ being a normalization constant.

In Fig.~\ref{fig:sketchRmatrix}, we present results for
\begin{align}\label{eq:Quench1}
    |f_{V}(E,\omega)|^2=\frac{1}{\sqrt{2\pi \sigma_\omega^2}}\exp(\frac{-\omega^2}{2\sigma_\omega^2}).
\end{align}
After the quench, $\braket{H_0(t)^2}$ shows an initial growth, followed by quick convergence to a steady state at long times. 

The SJA is accurate on all time-scales. Consider first time-scales $Jt$ of order one, where $\langle H_0^2\rangle_H$ rises from its initial value. All approximations are good in this regime. This serves as a consistency check for the SJA method, as the structure of its solution resembles second-order time-dependent perturbation theory at short times. At longer times, all approximations converge to a steady state. However, the results of TDPT and first-order SJA do not agree with the steady-state results of the exact dynamics. The agreement improves with higher-order SJA. 

\begin{figure*}[t!]
		\centering
		\includegraphics[width=1\textwidth]{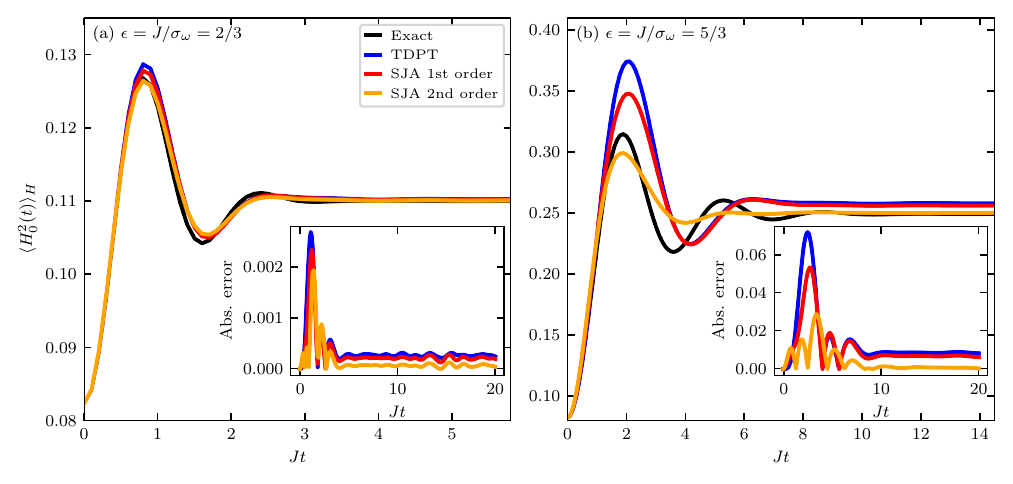}
		\caption{Evolution of $\braket{H_0^2}_H$ under a quench Eq.~\eqref{eq:Quench2}, $\sigma_\omega=0.3$, $\omega_0=0.7$, $N=2048$. Comparison of exact time evolution (black), with time-dependent perturbation theory (TDPT, blue), and the statistical Jacobi approximation to different orders (SJA, red, and orange). (a) $\epsilon=J/\sigma_\omega=2/3$, (b)~ $\epsilon=J/\sigma_\omega=5/3$. 
        Inset: the absolute error$|\braket{H_0^2}_{H,\text{approx}}-\braket{H_0^2}_{H,\text{exact}}|$.
        The data are averaged over 10 random realizations. While the oscillations at short times are not accurately captured by any approximate method for strong perturbations, the SJA captures the long-time dynamics in that case.
        } 
		\label{fig:sketchRmatrix2}
	\end{figure*}
    
As a next test, we consider another random matrix model, where we expect larger deviations between TDPT and SJA. As emphasized in Sec.~\ref{sec:flow_sol} and Eq.~\eqref{eq:fJac}, the difference between TDPT and first-order SJA is based on replacing $f_V(E,\omega)$ by $f_{\text{Jac}}(E,\omega)$. It is therefore instructive to consider a perturbation where these spectral functions differ significantly.

Such a case was already explored in Ref.~\cite{Long2023Beyond}, where the spectral function  of the perturbation
\begin{multline}\label{eq:Quench2}
    |f_{V}(E,\omega)|^2=\frac{1}{2\sqrt{2\pi \sigma_\omega^2}}\left[\exp(\frac{-(\omega-\omega_0)^2}{2\sigma_\omega^2})\right.\\
    \left.+\exp(\frac{-(\omega+\omega_0)^2}{2\sigma_\omega^2}) 
    \right].
\end{multline}
was chosen. $f_V(E,\omega)$ has two peaks at $\omega=\pm \omega_0$.
As it was shown in Ref.~\cite{Long2023Beyond}, there are significant deviations between $f_V(E,\omega)$ and $f_{\mathrm{Jac}}(E,\omega)$ around $\omega=0$. As the spectral function at small $\omega$ determines long-time dynamics, we therefore expect differences for the steady state at long times. 

The results for the perturbation in Eq.~\eqref{eq:Quench2} are shown in Fig.~\ref{fig:sketchRmatrix2}. Results are presented for small perturbations, specifically $\omega_0=0.7$, $\epsilon=J/\sigma_\omega=2/3$ (left), and larger perturbations $\epsilon=5/3$ (right). 
For the small perturbations $\epsilon=2/3$, all approximations show good agreement with the exact numerics at all timescales.
For $\epsilon=5/3$, none of the approximations capture the oscillations at intermediate times accurately, although the SJA solutions are closer to the exact dynamics.
At long times, second-order SJA is more accurate than TDPT, capturing the long-time saturation value within an absolute error of less than $0.02$.

It is important to note that for $\epsilon>1$, the separation of scales in the rescaled expression [Eq.~\eqref{eqn:flowBshorter}] does not hold. As a result, the iterative scheme is unjustified, potentially leading to significant deviations from the exact solution of the SJA flow equations at higher orders in the asymptotic expansion. It remains an open question whether the deviations between SJA and the exact dynamics at large $\epsilon$ are a feature of the iterative procedure or SJA itself.

\subsection{Spin-chain model}\label{subsec:spinmodel}
\begin{figure*}[t!]
		\centering
        \includegraphics[width=1\textwidth]{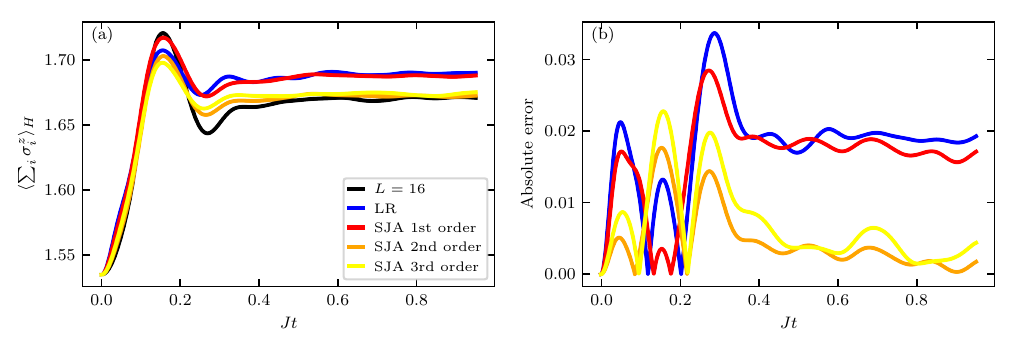}
		\caption{Evolution of $\braket{Z(t)}_H$ under a quench Eq.~\eqref{eq:pert}, starting from the mixed-field Ising Hamiltonian~Eq.~\eqref{eq:struct}; $J=0.1$, $L=16$. (a)~Comparison of exact time evolution (black), with linear response (LR, blue), and the statistical Jacobi approximation to different orders (SJA, red, orange and yellow).
        (b) Absolute error $|\braket{Z}_{H,\text{approx}}-\braket{Z}_{H,\text{exact}}|$.
        The initial growth is captured to some degree by all approximations. The higher-order approximations of SJA capture the steady state value more accurately.
        } 
		\label{fig:sketchstructure}
	\end{figure*}
Remarkably, the SJA outperforms the corresponding lowest-order
perturbative result to an even greater extent in structured models, as compared to random-matrix models.

Specifically, we test the SJA solution in a one-dimensional spin chain. This benchmark is important because the derivation of the flow equations neglects cross-correlations between matrix elements during the Jacobi algorithm~[Sec.~\ref{subsec:flow_deriv}]. While such assumptions are natural in the random matrix model, they require testing in models with more structure, such as spin-chain models. 

For our analysis, we choose the mixed-field Ising model
\begin{align}\label{eq:struct}
    H_0=\sum_i \sigma^z_{i}\sigma^z_{i+1}+ g \sigma^x_i+ h \sigma^z_i,
\end{align}
with field strengths $g=0.9045$ and $h=0.809$, and periodic boundary conditions. The model is well known to thermalize rapidly~\cite{Kim2013Ballistic}. 
As a perturbation, we take 
\begin{align}\label{eq:pert}
    J\,V=J \sum_i \sigma^x_i\sigma^x_{i+1}-\sigma^y_{i}\sigma^y_{i+1}.
\end{align}
For the following numerics, we restrict to the zero-momentum sector.
The initial density matrix is given by
\begin{align}
    \rho=\frac{1}{\mathcal{N}}\sum_{|E_i|<0.5 L}\ketbra{i}{i},
\end{align}
with $\mathcal{N}$ being the normalization factor.

We compute the dynamics of the total magnetization $Z$ defined in Eq.~\eqref{eq:Z}.

The results are shown in Fig.~\ref{fig:sketchstructure} for $J=0.1$.
All approximations reproduce the short-time growth.
The height of the first peak in the oscillations and the long-time saturation value are however only reproduced by second-order SJA. 
As before, second-order SJA reduces the absolute error at long times significantly in comparison to linear response.

\subsection{Results for autocorrelators}

The SJA can also be used to predict the auto-correlator of \(A\) in a thermal state of \(H\) from the solution of Eq.~\eqref{Eq:flowAutocor}. Time evolution is also generated by \(H\). This contrasts to quench dynamics, where, while evolution is generated by \(H\), the initial state was taken to be a stationary state of \(H_0\).

\begin{figure*}[t!]
		\centering
        \includegraphics[width=1\textwidth]{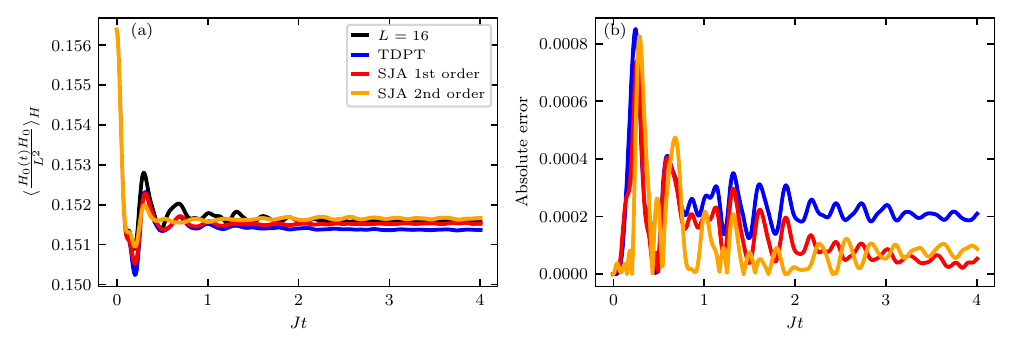}
		\caption{Evolution of the infinite temperature auto-correlator $\braket{H_0H_0(t)}_H$ where $H$ is the perturbed mixed-field Ising Hamiltonian~Eq.~\eqref{eq:pert} with $J=0.2$, $L=16$. (a)~Comparison of exact time evolution (black), with second-order time-dependent perturbation theory (TDPT, blue), and the statistical Jacobi approximation to different orders (SJA, red and orange).
        (b) Absolute error $|\braket{H_0(t)H_0}_{H,\mathrm{approx}}
-\braket{H_0(t)H_0}_{H,\mathrm{exact}}|/L^2$.
        The initial decay is captured by all approximations, while the SJA approximates the steady state value better than TDPT. 
        } 
		\label{fig:sketchAutocor}
	\end{figure*}

Fig.~\ref{fig:sketchAutocor} plots the infinite temperature auto-correlator $\langle H_0(t)H_0\rangle_H$ for the different approximations in the mixed-field Ising model with $J=0.2$. The initial decay is reproduced by all approximations. Again, as before, first- and second-order SJA capture the long-time saturation value of the autocorrelator better than TDPT.

\subsection{Finite-size scaling}\label{sec:Finitesizescaling}
\begin{figure*}[t!]
		\centering
        \includegraphics[width=1\textwidth]{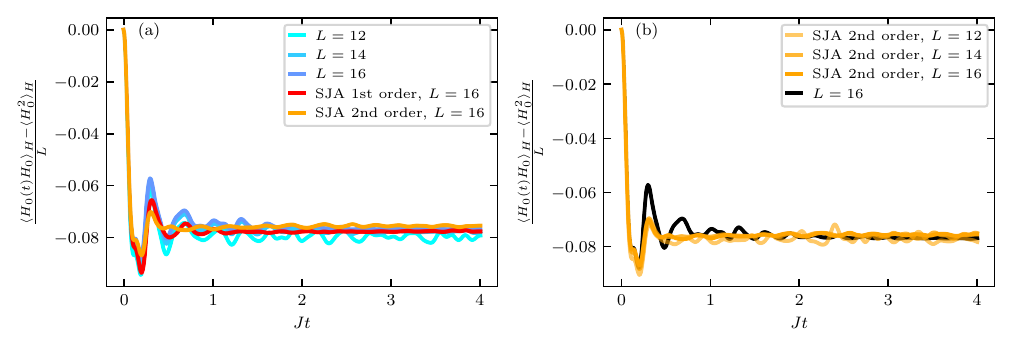}
		\caption{Evolution of the infinite temperature auto-correlator $\langle H_0(t) H_0 -H_0^2\rangle_H/L$  in the perturbed mixed-field Ising Hamiltonian~Eq.~\eqref{eq:pert} with $J=0.2$, and \(L\) varying from \(12\) to \(16\). (a)~Comparison of exact results for $L=12$, $L=14$ and $L=16$ with first and second-order SJA~($L=16$).
        (b) Comparison of second-order SJA results for $L=12$, $L=14$ and $L=16$ with exact data for $L=16$. The predictions from SJA are stable with increasing the system size and show less finite-size fluctuations as compared to the exact data. 
        } 
		\label{fig:sketchFinitesizescsaling}
	\end{figure*}
Finally, we show that the SJA predictions for the autocorrelator $\langle H_0(t) H_0 -H_0^2\rangle_H/L$ (an intensive quantity) exhibit only weak system-size dependence over the system sizes considered. This provides evidence that the iterative solutions are in the thermodynamic limit, and the numerical statistical input from the Jacobi algorithm, $\rho_{\mathrm{dec}}$, and the initial form factors exhibit stable features upon increasing the system size. Thus, the SJA can predict the dynamical behavior of large systems using $\rho_\mathrm{dec}$ and the initial form factors computed at small system sizes.

The results are shown in Fig.~\ref{fig:sketchFinitesizescsaling}. Fig.~\ref{fig:sketchFinitesizescsaling}~(a) compares numerically exact data at different system sizes. In Fig.~\ref{fig:sketchFinitesizescsaling}~(b), we compare SJA predictions using $\rho_{\text{dec}}(w,E,E^\prime)$ and the initial form factors obtained from different system sizes $L=12$ to $L=16$.
The time evolution of the exact dynamics and the SJA predictions in both figures show system-size dependent fluctuations for intermediate and long times. However, the fluctuations in the SJA predictions are suppressed in comparison to fluctuations in the exact data.

\section{Discussion}
	\label{sec:disc}
In this work, we have taken the first steps toward making the ETH a predictive theory for dynamical response functions in well-thermalizing systems. 

Our procedure, dubbed the Statistical Jacobi Approximation (SJA) requires, as input, an initial state $\rho$, an observable of interest \(A\), and various form factors in the ETH ansatz for $A$ and $\rho$ in the eigenbasis of the unperturbed Hamiltonian $H_0$. These form factors determine response functions with respect to the Hamiltonian $H_0$. It also requires a statistical description of the Jacobi algorithm, which rotates the eigenbasis of $H_0$ to that of a perturbed Hamiltonian \( H =  H_0 + J V \) through a series of two-level rotations.

The output of the SJA is a set of response functions with respect to the perturbed Hamiltonian $H$, specifically \( \braket{A(t)}_H \) upon quenching from $H_0$ to $H$, and auto-correlators in the thermal ensemble of $H$. The SJA assumes that the ETH holds after sufficiently many rotations of the Jacobi algorithm and derives a flow equation for the form factors in the ETH ansatz. Solutions to the flow equations predict the desired response functions. Our approximate solutions to these flow equations compare well to numerically exact solutions in random matrix models and in one-dimensional spin chains.

As a numerical technique at fixed system size, the Jacobi algorithm is not competitive with state-of-the-art exact diagonalization or tensor-network based approaches~\cite{Vidal2004Efficient,White2004Real-Time,Hageman2011Time,Schollw_ck_2011}. However, the SJA is not as plagued by finite-size effects, for two reasons. 

First, the flow equations hold in the thermodynamic limit. Future work could apply sophisticated numerical techniques to directly solve them. Next, the statistical distribution of the Jacobi rotations [Eq.~\eqref{eq:rhodecdefjoint}], which is computed numerically exactly at small system sizes, is expected to be stable with increasing system size after rescaling. The stability follows from the Jacobi algorithm's organization of the rotation by scale $w$. Larger values of $w$ are less affected by finite-size effects. Happily, the large $w$ part of the distribution also determines the largest corrections to response functions in well-thermalizing systems.

More broadly, the SJA method disentangles the different contributions to the dynamics according to the timescales at which they appear. We thus believe that the SJA will be generally useful in problems with a wide range of time scales.

Part of the appeal of the SJA is that the path to generalization and computing other quantities of interest is clear: take the generalized ETH ansatz~\cite{Foini2019Eigenstate,Pappalardi2022Eigenstate} with respect to the Hamiltonian $H_0$, compute the flow equations for all the relevant form factors as in Sec.~\ref{sec:flow}, and obtain the desired response functions with respect to the Hamiltonian $H$. The out-of-time ordered correlators may be a good future target~\cite{Chan2019Eigenstate,Foini2019Eigenstate,Pappalardi2022Eigenstate,Hahn2024Eigenstate}.

Current applications of the SJA take the distribution of matrix elements decimated by the Jacobi algorithm, \(\rho_{\mathrm{dec}}\), as an input. A powerful extension of the framework would be to predict \(\rho_{\mathrm{dec}}\) directly from the statistical description of the perturbation \(V\) in the Jacobi basis. Since the perturbation \(V\) is itself affected by the Jacobi rotations, we anticipate that a flow equation for \(\rho_{\mathrm{dec}}\) will be non-linear~\cite{Vanoni2026resonance}. 

In upcoming work~\cite{Hahn2025Floquet}, we extend the SJA framework to periodically driven Floquet systems. The external drive adds an extra dimension to the form factors, namely the harmonic of the drive frequency. We use this extended framework to investigate the physics of heating~\cite{Abanin_2017,Abanin2017Effective,Morningstar2023Universality}, and the crossover between heating and non-heating regimes in mesoscopic systems~\cite{Bukov2016Heating,Morningstar2023Universality}.

Finally, we note that the form factors that are predicted by the flow equations can be also defined in classical systems, as Fourier transforms of dynamical correlation functions. It is an open question if the SJA applies to classical systems, and if it produces useful classical-quantum correspondences in many-body systems.

\acknowledgements
    \label{sec:ack}
    We thank Tobias Helbig, Stefan Kehrein, Chris Laumann, and Silvia Pappalardi for useful discussions. We thank Carlo Vanoni for useful comments on the manuscript.
    This work was supported by: the European Union (ERC, QuSimCtrl, 101113633) (DH and MB);
the Leverhulme Trust [Grant No. LIP-2020-014]~(DH); the Laboratory for Physical Sciences (through their support of the Condensed Matter Theory Center at the University of Maryland) (DML), NSF Grant No. DMR-1752759 (AC and DML), a Stanford Q-FARM Bloch postdoctoral fellowship (DML), the US Department of Energy,
Office of Science Award No. DE-SC0019380 (DML), and a Packard Fellowship in Science and Engineering (DML, PI: Vedika Khemani). AC thanks the Max Planck Institute for the Physics of Complex Systems for its hospitality.
Views and opinions expressed are however those of the authors only and do not necessarily reflect those of the European Union or the European Research Council Executive Agency. Neither the European Union nor the granting authority can be held responsible for them.
Numerical simulations were performed on the MPIPKS HPC cluster.

\bibliography{fidelity_Jacobi.bib,operator.bib}
	
	\appendix
\begin{widetext}
   \section{The statistical Jacobi approximation and linear response}\label{sec:Linear repsonse}

    We mentioned in Sec.~\ref{sec:flow_sol} that the  terms linear in $\sin \eta/2$ are related to the linear response function.
    To see this, consider the first two terms in the second line of Eq.~\eqref{eq:flowequationstart2}. 
    At the lowest order, Jacobi decimates elements of $H$ one by one without affecting the other matrix elements. Using
$w_n e^{-\I\phi_n}=J V_{ab}$ and
$\sin(\eta_n/2)\simeq w_n/(E_{a}-E_{b})$, we obtain
\begin{equation}\label{eq:Linearterm}
e^{-\I\phi_n}\sin\frac{\eta_n}{2}
\left(\rho^{(n)}_{bb}A^{(n)}_{ba}-\rho^{(n)}_{aa}A^{(n)}_{ba}\right)
\simeq
J\frac{V_{ab}}{E_{a}-E_{b}}
\left(\rho^{(n)}_{bb}A^{(n)}_{ba}-\rho^{(n)}_{aa}A^{(n)}_{ba}\right).
\end{equation}

Compare this expression with the Kubo formula
\begin{align}
\begin{split}
	\braket{A(t)}_H-\braket{A(0)}_H=\int \dd\omega\,\chi(\omega)
\left(\e^{\I\omega t}-1\right). 
\end{split}
\end{align}
 $\chi(\omega)$ is for $\omega\neq 0$ given by
\begin{align}\label{eq:spectrum}
	\chi(\omega)=J\sum_{a_0,b_0}
\delta\left(\omega-(E_{b_0}-E_{a_0})\right)
\frac{\rho_{a_0a_0}V_{a_0b_0}A_{b_0a_0}-
\rho_{b_0b_0}V_{a_0b_0}A_{b_0a_0}}{E_{b_0}-E_{a_0}}.
\end{align}
The summands in Eq.~\eqref{eq:spectrum} agree with Eq.~\eqref{eq:LRcomparison}.

\section{Second-order perturbation theory}\label{app:Second order perturbation theory}
In the following section, we recapitulate results of second-order perturbation theory. We only keep results up to the second order in the perturbation $V$. 

Consider the expectation value
\begin{align}
    \braket{A(t)}_H=\Tr[\rho(t)A].
\end{align}
As in the main text, we consider the time evolution under a quench $H=H_0+JV\Theta(t)$, and $\rho(0)$ and $A$ are diagonal in an eigenbasis of $H_0$.
$\rho(t)$ is given by
\begin{align}\label{eq:orig}
    \rho(t)=\e^{-\I H_0 t}U(t) \rho(0)U^\dagger(t) \e^{+\I H_0t}.
\end{align}
with $U(t)$ given by
\begin{align}
\begin{split}
    &U(t)=1-\I \int_0^t \e^{+ \I H_0t'}J V \e^{-\I H_0t'}\,\mathrm{d}t'-\int_0^t \e^{+\I H_0t'}J V \e^{-\I H_0t'}\,\mathrm{d}t'\int_0^{t'} \,\mathrm{d}t'' \e^{+\I H_0t''}J V \e^{-\I H_0t''}+\mathcal{O}(J^3)
\end{split}
\end{align}
In the eigenbasis of $H_0$, this expression is given by:
\begin{align}
\begin{split}
    U(t)&=1- J\sum_{n_0,m_0} \frac{\e^{\I(E_{n_0}-E_{m_0})t}-1}{E_{n_0}-E_{m_0}}\braket{n_0|V|m_0}\ket{n_0}\bra{m_0}\\&+
    J^2\sum_{n_0,m_0,k_0} \left(\frac{\e^{\I(E_{n_0}-E_{m_0})t}-1}{(E_{n_0}-E_{m_0})(E_{k_0}-E_{m_0})}\right.\left.-\frac{\e^{\I(E_{n_0}-E_{k_0})t}-1}{(E_{n_0}-E_{k_0})(E_{k_0}-E_{m_0})}\right)\braket{n_0|V|k_0}\braket{k_0|V|m_0}\ket{n_0}\bra{m_0}+\mathcal{O}(J^3)
\end{split}
\end{align}
This expression can be reinserted into Eq.~\eqref{eq:orig}.
The first-order correction in $J$ reproduces Kubo's formula
\begin{align}
    -\I \int_0^t \Tr([A,\e^{\I H_0t'}J V\e^{-\I H_0t'}]\rho(0))\,\mathrm{d}t'
\end{align}
Since $A$ is chosen to be diagonal in the eigenbasis of $H_0$ and thus commutes with $\rho(0)$ and $H_0$, the first order vanishes in that case.

For second order, we obtain:
\begin{equation}
    \braket{A(t)}_H=\braket{A(0)}_H+J^2\sum_{n_0,m_0}\frac{|\braket{n_0|V|m_0}|^2}{(E_{n_0}-E_{m_0})^2}\left(\braket{n_0|A|n_0}-\braket{m_0|A|m_0}\right)\left(\braket{n_0|\rho|n_0}-\braket{m_0|\rho|m_0}\right) (\e^{-\I(E_{n_0}-E_{m_0})t}-1)
\end{equation}
With the definitions of form factors Eq.~\eqref{eq:definitions}, Eq.~\eqref{eq:definitipons3}, and Eq.~\eqref{eq:deffv}, we obtain
\begin{equation}
\braket{A(t)}_H-\braket{A(0)}_H
=\int \rd E\, \rd\omega\,
J^2\frac{|f_V(E,\omega)|^2}{\omega^2}
\left[p(w_0,E)-\frac{\nu(E)}{\nu(E+\omega)} p(w_0,E+\omega)\right]
\left[A(w_0,E)-A(w_0,E+\omega)\right]\left(\e^{\I\omega t}-1\right).
\end{equation}
\end{widetext}

\end{document}